\documentclass[reqno,12pt]{amsart}

\usepackage{amsfonts}

\usepackage[comma, authoryear]{natbib}
\usepackage{enumerate}
\usepackage{epsfig}
\usepackage{graphicx}
\usepackage{subfigure}
\usepackage{caption}
\usepackage[utf8]{inputenc}
\usepackage{amsmath}
\usepackage{amssymb}
\usepackage{mathrsfs}
\usepackage{float}

\numberwithin{equation}{section}

\newtheorem{thm}{Theorem}

\newtheorem{RM}{Remark}

\newtheorem{ex}{Example}

\begin{document}
\title[New Simulation and Pricing]{Explicit Heston Solutions and Stochastic Approximation for Path-dependent Option Pricing}

\author[M. Kouritzin]{By Michael A.  Kouritzin}
\curraddr{Department of Mathematical and Statistical Sciences\\
	University of Alberta  \\
	Edmonton (Alberta)\\
	Canada T6G 2G1}
\email{michaelk@ualberta.ca
	\newline\indent {\it URL:} http://www.math.ualberta.ca/Kouritzin\_M.html}

\thanks{Partial funding in support of this work
	was provided by NSERC Discovery Grant 203089.}
\renewcommand{\subjclassname}{\textup{2010} Mathematics Subject Classification}
\dedicatory{\large University of Alberta} \maketitle

\begin{abstract}
New simulation approaches to evaluating path-dependent options without matrix inversion issues nor
Euler bias are evaluated.
They employ three main contributions:
(1) stochastic approximation replaces regression in the LSM
algorithm;
(2) explicit weak solutions to 
stochastic differential equations are developed and applied
to Heston model simulation; and
(3) importance sampling expands these explicit solutions.
The approach complements Heston (1993) and 
Broadie \& Kaya (2006) by handling the case of path-dependence in the option's execution strategy.
Numeric comparison against standard Monte Carlo methods
demonstrate up to two orders of magnitude speed improvement.
The general ideas will extend beyond the important Heston setting.
\end{abstract}
	

\section{Introduction}\label{intro}
The optimal pricing of American and other path-dependent options for multiple factor
models remains problematic.
Traditionally, finite difference methods have been used (see e.g. 
Schwartz 1977, Wilmott et. al. 1995) to solve
the corresponding partial differential equation.
However, they can be computationally expensive when the model has
multiple factors and also complicated
to adapt when the model has jumps.
This has led to the development and use of Monte Carlo based pricing methods 
(see e.g. 
Boyle 1977, Duffie \& Glynn 1995, Boyle et. al. 1997, Carriere 1996),
for which one needs simulation.
A most successful simulation method for Monte Carlo multi-factor, path-dependent
option pricing is the LSM algorithm developed by \cite{LoSc} and further
analyzed by \cite{ClPr}.
As usual, they approximate American (and other continuously-executable) options discretely, implementing and
analyzing the resulting Bermuda-style options.
However, there are problems.

\subsection{Motivational Problem}
Suppose we wanted to price an American (really Bermudan) Put option based upon the Heston 
model (see (\ref{Heston}) to follow) with Heston and option parameters:
$\nu={8.1\kappa^2}/{4},\mu=0.0319,\rho=-0.7,\varrho=6.21,\kappa=0.2 $, option duration $T=50$, 
initial price $S_0=100$, initial volatility $V_0=0.102$, and the strike price $K=100$.
The fair price of this option will turn out to be \$$7.9426$.
However, if we use the LSM algorithm and Monte Carlo simulation with Euler or the Implicit Milstein 
approximations of
\cite{KaJa}, then
the best we can get on an inexpensive contemporary computer is \$$7.371$ for as we try
to go beyond that the algorithm fails numerically, producing smaller values while
taking longer times to compute.
(Throughout this paper references to the Milstein method will always mean the Implicit Milstein method
proposed by Kahl and J\"{a}ckel as the normal Milstein method does not perform well.)
Our goals herein are to get around the numeric least squares regression problems of
the LSM algorithm and the slow, biased nature of the Euler and Milstein simulation methods.
We do this by explicit weak solutions and stochastic approximation.
The result will be a \emph{three order of magnitude speed improvement} in simulation
and a \emph{two order of magnitude speed improvement} in path-dependent option pricing.

\subsection{The LSM/Simulation Setting}
Suppose there is a complete
filtered (risk-neutral) probability space $(\Omega,\mathcal F,\{\mathcal F_t\}_{t=0}^T,P)$
supporting a Markov chain $\{(S_t,V_t)\}_{t=0}^T$ with state space $D=D_S\times D_V$, representing 
the observable and hidden components of the asset state (like price and
volatility), as well as the (discounted) adapted payoff $Z_t\ge0$ received for executing the 
option at time $t\in[0,T]$.
(In many settings there are multiple risk-neutral measures and one
is chosen by calibrating to model market data.
In the Heston case, the volatility component causes the non-uniqueness
and should be calibrated using e.g.\ real option prices.  
We assume throughout that this has been done.)
Then, the option-pricing objective is to compute $\sup_{\tau_0\in\mathcal T_{0,T}}E[Z_{\tau_0}]$,
where $\mathcal T_{t,T}$ denotes the collection of stopping times with values in $\{t,t+1,...,T\}$.
Using dynamic programming and following \cite{ClPr}, one finds a best $\tau_0\in\mathcal T_{0,T}$
by working backwards according to
\begin{equation}
\left\{\begin{array}{lll}\tau_T&=&T\\
	\tau_t&=&t1_{\{Z_t\ge E[Z_{\tau_{t+1}}|\mathcal F_t]\}\cap\{Z_t>0\}}+\tau_{t+1}1_{\{Z_t< E[Z_{\tau_{t+1}}|\mathcal F_t]\}\cup\{Z_t=0\}}
\ \forall\ t<T\end{array}\right.\!.\ \ 	
\end{equation}
Typically, $E[Z_{\tau_{t+1}}|\mathcal F_t]>0$ so $\cap\{Z_t>0\}$ and $\cup\{Z_t=0\}$ do not effect the recursion.

Now, assume:
\begin{description}
\item[Total] there are measurable real-valued functions $(f_t)_{t=0}^T$ 
and $(e_k)_{k=1}^\infty$ on $D$ such that $E[Z_{\tau_t}|\mathcal F_t]=f_t(S_t,V_t)$ for all $t=0,...,T$ and $\{e_k(S_t,V_t)\}_{k=1}^\infty$ is total 
on $L^2(\sigma(S_t,V_t),1_{\{Z_t>0\}}dP)$ for all $t=1,...,T-1$.
\end{description}
{A subset of a Hilbert space is total if its span is the entire space.}

Following \cite{LoSc} to create the $(e_k)_{k=1}^\infty$, we often start with bases functions 
$(e^S_{k})_{k=1}^\infty$, $(e^V_{k})_{k=1}^\infty$ on 
$L^2(D_S)$, $L^2(D_V)$ respectively and
let $(e_k(s,v))_{k=1}^\infty$ be some ordering of
$\{e^S_{k_1}(s)e^V_{k_2}(v)\}_{k_1,k_2=1}^\infty$.

The key idea in the LSM algorithm is to estimate the conditional expectations
$E[Z_{\tau_{t}}|\mathcal F_t]$ (by first estimating
$E[Z_{\tau_{t+1}}|\mathcal F_t]$)
from the cross-sectional data using projection $P^J_t$ onto the closed linear span of $\{e_k(S_t,V_t)\}_{k=1}^J$ and 
least-squares regression.
Indeed, \cite[Theorem 3.1]{ClPr} show that
\begin{equation}\label{L2Converge}
\lim_{J\rightarrow\infty}E[Z_{\tau_t^J}|\mathcal F_t]=E[Z_{\tau_t}|\mathcal F_t]
\end{equation}
in $L^2$ for all $t\in\{0,...,T\}$, where 
\begin{equation}
\left\{\begin{array}{lll}\tau_T^J&=&T\\
	\tau_t^J&=&t1_{\{Z_t\ge P^J_t[Z_{\tau^J_{t+1}}]\}\cap\{Z_t>0\}}+\tau_{t+1}^J1_{\{Z_t< P^J_t[Z_{\tau^{J}_{t+1}}]\}\cup\{Z_t=0\}}
\ \forall\ t<T\end{array}\right..	
\end{equation}
Then, letting $e^J=(e_1,...,e_J)'$ (where $a'$ denotes transpose of
vector or matrix $a$) and assuming
\begin{description}
	\item[Non-singular] $E[e^J(S_t,V_t)(e^J(S_t,V_t))'1_{\{Z_t>0\}}]$ is positive definite,
\end{description}
\cite{LoSc} recognize that the $\alpha^J_t$ in $P^J_t[Z_{\tau^J_{t+1}}]=\alpha^J_t\cdot e^J(S_t,V_t) $
is $\alpha^J_t=E[e^J(S_t,V_t)(e^J(S_t,V_t))'1_{\{Z_t>0\}}]^{-1}E[Z_{\tau^J_{t+1}}e^J(S_t,V_t)1_{\{Z_t>0\}}]$ i.e.\ the solution to
\begin{equation}\label{L2Opt}
	\min_{\alpha^J}E[|Z_{\tau^J_{t+1}}-\alpha^J\cdot e^J(S_t,V_t)|^21_{\{Z_t>0\}}],
\end{equation}
which they solve by Monte Carlo simulation and linear regression:
Let $\{(S^j,V^j,Z^j)\}_{j=1}^N$ be i.i.d.\ copies of $(S,V,Z)$
and the ${\tau^{J,j}_{t+1}}$ satisfy 
\begin{equation}\label{tauJjdef}
\!\left\{\begin{array}{lll}\tau_T^{J,j}&=&T\\
	\tau_t^{J,j}&=&t1_{\{Z^j_t\ge P^{J}_t[Z^j_{\tau^{J,j}_{t+1}}]\}\cap\{Z^j_t>0\}}+\tau_{t+1}^{J,j}1_{\{Z^j_t< P^{J}_t[Z^j_{\tau^{J,j}_{t+1}}]\}\cup\{Z^j_t=0\}}
\ \forall\ t<T\end{array}\right.\!.	
\end{equation}
Then, their least squares estimate is $\alpha^{J,N}_t=(A^N_t)^{-1}b^N_t$ with 
\begin{equation}
A^N_t=\frac1N \sum_{j=1}^N e^J(S^j_t,V^j_t)e^J(S^j_t,V^j_t)' 1_{Z^j_t>0},\ \ 
b^N_t=\frac1N \sum_{j=1}^N Z^j_{\tau^{J,j}_{t+1}}e^J(S^j_t,V^j_t) 1_{Z^j_t>0}.
\end{equation}
Notice that $\tau_t^{J,j}$ depends on $P^{J}_t[Z^j_{\tau^{J,j}_{t+1}}]$ which depends upon $\alpha^{J,N}_t$ which in
turn depends upon $\tau^{J,j}_{t+1}$, meaning we must construct these objects in
reverse time and at each time compute $\alpha^{J,N}_t$ prior to $\tau_t^{J,j}$.

\subsection{Weaknesses of Current Methods}
The LSM algorithm has a weakness:
The regression requires inverting a (generally) dense $J\times J$ matrix $A^N_t$
with random coefficients,
which becomes ill-conditioned as
the number of factors in the model or the desired accuracy (and consequently the number of bases
functions $J$ required) increases.
Many examples given in \cite{LoSc} have features that may allow a lower
number of basis functions:
Shorter durations facilitate a smaller $J$ because
there are fewer possible execution times to choose from in the Bermudian approximations.
Single factor models make projection one dimensional, which generally facilitates
better approximation with fewer functions versus higher dimensional projection.
American put options with strike price $K$ effectively restrict $S$ to $[0,K]$ or less,
which also makes the projection ``easier".
The need for lower accuracy reduces the required $J$ as it becomes acceptable
to get more of the optimal stopping possibilities wrong.
Not all problems have these features.
Yet, the most bases functions used in \cite{LoSc} was $26$.  
In some examples below, $J$ will need to be much larger, making matrix
inversion problematic.
Fortunately, there is a stochastic approximation alternative and it is also 
faster than regression.
This is the first main contribution of this paper.

The other major problems with the simulation approach to path-dependent option
pricing are computation time and bias.
The famous geometric Brownian motion (GBM) model, utilized in the classical Black-Scholes
option pricing formula (see Black \& Scholes  1973, Merton 1973),
has constant volatility and follows
the linear stochastic differential equation (SDE)
\begin{equation}
	\label{GBM} dS_{t}=\mu S_t\, dt+\kappa S_t \,dB_{t},
\end{equation}
where $B$ is a standard Brownian motion and $\mu,\ \kappa$ are the drift and volatility parameters.
It is well known that the GBM model is overly simplistic, results
in unnatural phenomena like the volatility smile commonly observed in market
option prices (see \cite{JaRu} for a detailed survey) and should be replaced
by stochastic volatility (SV) models with two components:
price $S$ and stochastic variance $V$ (or volatility $V^\frac12$) that
replaces the constant $\kappa$ in the GBM model.

\cite{He} introduced a stochastic volatility model with closed
form European-call-option prices for stock, bond
and foreign currency spot prices.
Let $B$, $\beta$ to be (scalar) independent standard Brownian motions.
Then, the {Heston} model is:
\begin{equation}\label{Heston}
d\left(\begin{array}{c}S_{t}\\V_{t}\end{array}\right)
=\left(\begin{array}{c}\mu S_{t}\\\nu-\varrho V_{t}\end{array}\right)dt+\left(\begin{array}{cc}\sqrt{1-\rho^2}S_{t}V_{t}^{\frac{1}{2}}&\rho S_{t}V_{t}^{\frac{1}{2}}\\0&\kappa V_{t}^{\frac{1}{2}}\end{array}\right)\left(\begin{array}{c}dB_{t}\\d\beta_{t}\end{array}\right),
\end{equation}
with parameters $\mu\in\mathbb R$, $\rho\in[-1,1]$ and $\nu,\varrho,\kappa>0$.
The volatility component is just the Cox-Ingersoll-Ross (CIR) model.
The volatility can hit $0$ when $\nu<{\kappa^2}/2$ and can still approach
$0$ when the Feller condition $\nu\ge{\kappa^2}/2$ holds.
From a financial perspective, hitting zero would imply randomness coming out of the
price, which is not common, so we generally have $\nu$ larger than ${\kappa^2}/2$.
An important feature of the Heston model is that it allows arbitrary
correlation $\rho\in [-1,1]$ between volatility and spot asset returns.
Indeed, $\rho$ is often negative in financial markets (see e.g.\ 
Fouque et. al. 2000 p. 41, Yu 2005).
The Heston model can be used to explain and correct for 
skewness and strike price bias and to outperform other popular SV models on real data
(see Kouritzin 2015
for the later).
Broadie \& Kaya (2006) developed an exact (without bias)
simulation method for the Heston model to price
options with at most weak path dependence.
This paper addresses the remaining significant difficulty,
effectively
pricing path-dependent Heston options including the American and Asian options.
Herein, the Heston model stochastic differential equations (SDEs)
are solved explicitly in weak form and these solutions are used to
price options and do Monte Carlo simulations.

The Euler-Maruyama  and Milstein simulation methods have obvious problems for the Heston model:
1) While the process itself is nonnegative, the discretization may try producing negative values
causing evaluation issues when square rooted.
2) The rate of convergence to the actual diffusion is slow.
In fact, \citet{BroadieKaya:2006} did a nice job of demonstrating
the bias problem of these methods even when the computations are appropriately balanced 
in the sense of \citet{DufGly}.
3) The computation time is large, making real-time application more
difficult for higher-volume, rapidly-traded equities.
For example, the use of Euler-Maruyama and Milstein methods 
made real-time application (versus back data study) impossible in \citet{KoMF}.
Hence, exact simulation as in \citet{BroadieKaya:2006},
where Heston model specifics are used to avoid bias and increase speed, is desired.
Unfortunately, this type of exactness (in terms of distribution transforms)
is not amenable to valuing American, Asian and other heavily-path-dependent
options.
Herein, we introduce explicit weak solutions
to the Heston SDEs, our most significant contribution, which makes simulation and Monte Carlo path-dependent
option pricing relatively easy.
We introduce new pricing algorithms, give new theorems for explicit solutions,
develop new methods for finding explicit solutions and provide American and Asian
option pricing examples.

\subsection{Main Contributions}

The first, and most famous, Stochastic Approximation (SA) algorithms are the 
Robbins-Monro and Kiefer-Wolfowitz algorithms introduced respectively in \cite{RoMo51} and \cite{KiWo52}.
SA algorithms were initially applied to find roots and maxima of functions
defined in terms of expected value.
Since then, they have become important methods in statistics and engineering for
such things as parameter estimation in time series and channel
equalization in communications.
Engineers often consider SA algorithms as part of adaptive filtering but
many SA algorithms, like the so-called sign algorithm, are not linear.
\cite{Eweda} contains a comparison of three popular algorithms on a
communication channel problem while \cite{KoSa15} uses an SA algorithm
in linear observer design.
To the author's knowledge, this paper is the first use of SA in option pricing
and it differs from other applications in the sense that stochastic
approximation is applied across particles and not over time.

Explicit solution of the CIR model, which constitutes the volatility part
of the Heston model, has been known for twenty years, dating back to
(at least) \cite{Maghsoodi} in general and \cite{Kouritzin:2000} in a 
form similar to that used here.
Moreover, it is clear that the Heston price is a stochastic exponential
given the volatility but this stochastic exponential would still involve a function of $\int _{0}^{t}V_{s}^{\frac{1}{2}}d\beta_{s}$
with $\beta$ and $V$ being dependent so stochastic integral approximation would
seem to be required.
Separately, \citet{BroadieKaya:2006} give an exact (marginal) distributional description
of the Heston model and use that to simulate at a fixed time.
(Chapter 6 of \citet{JeYoCh} also provides a very nice overview of the CIR and Heston models
with both the explicit solution of the CIR model and the single-time-marginal distributional exactness of the
Heston model.)
However, none of these works give us an explicit solution for 
(both price and volatility components of) the
Heston model and as such do not provide an alternative means for
pathwise simulation.
The second main contribution of this paper is to show there is such
an explicit solution.
The price part of this solution under Condition (C) below has the form 
$\phi _{t}\left(\int _{0}^{t}V_{s}^{\frac{1}{2}}dB_{s},\int _{0}^{t}V_{s}ds,V_{t}\right)$
for some known $\phi$, where $B$ is independent of $V$.
This means $\int _{0}^{t}V_{s}^{\frac{1}{2}}dB_{s}$ is conditionally
Gaussian and there is no need for approximating stochastic integrals.
The volatility part basically comes from \cite{Maghsoodi}'s observation.
Together they produce an efficient means to simulate the Heston 
model, given as the explicit Heston method below.

When Condition (C) is not fulfilled, one uses a likelihood weight to convert
to a new set of parameters where it is fulfilled.
In this case, the weighted Heston algorithm below still gives an
explicit solution.
This conversion is our third contribution and is basically importance
sampling for the variance.
While importance sampling is used in many areas of statistics, our
third contribution is to use it to maintain explicit solutions and
produce a weighted particle method of option pricing.
We are in effect using sequential Monte Carlo methods in pathspace
option pricing.
The weighted particle sequential Monte Carlo method maintains
pathspace estimates as required for path-dependent option pricing.

\subsection{Layout}
The remainder of this paper is laid out as follows:
Our new algorithms and theoretical results are given in Section 2.
The first algorithm is a stochastic approximation variation of the LSM
algorithm.
The second algorithm is for simulating Heston SDEs.
It fits into the first algorithm when the Heston model is used and is based
upon our main theorems.
The first theorem gives basic explicit solutions that hold under a restriction
on the parameters of the Heston model.
The second result provides weak solutions when this restriction does not hold.
Section 3 compares our new Heston simulation algorithms to the Euler-Maruyama and Milstein 
simulation methods and shows a \emph{three order} of magnitude speed improvement
for the same accuracy.
Section 4 compares our new Heston simulation and SA algorithms to the LSM algorithm as 
well as the Euler-Maruyama and Milstein 
simulation methods on the American and Asian option pricing problems.
In particular, pricing of put, call and straddle
options are considered for the Heston model and the combined effect of
the new simulation and SA algorithms are shown to provide a \emph{two order} of magnitude 
improvement on pricing such options compared to the standard LSM/Euler or LSM/Milstein
approach.
Our conclusions are in Section 5 and our proofs are relegated to 
the appendix, which is Section 6.
There are quick unmotivated, \emph{guess-and-check} proofs of our theorem using It\^{o}'s formula.
However, our proofs are really our 
method of finding explicit (weak) solutions for financial models.
Hence, they could turn out to be the most important part of this work if they
provide a means of coming up with weak solutions to other financial models
as it is believed they will. 


\section{Algorithms and Results} \label{FinModWeak}

\subsection{Stochastic Approximation Pricing Algorithm}

Stochastic Approximation (SA) algorithms solve stochastic optimization problems
like the mean-square optimization problem (\ref{L2Opt}).
Our application is similar to the SA framework of \cite{Kouritzin:1996} and \cite{KoSa15}.
Suppose $\{(L^j,S^j,V^j,Z^j)\}_{j=1}^N$ are i.i.d.\ copies of $(L,S,V,Z)$, where
$S,V,Z$ are as in the introduction and
$L$ is some likelihood, i.e.\ a non-negative martingale and satisfying $E[L_t]=1$
for all $t$.
$L$'s purpose is to reweight $(S,V,Z)$ so they have the correct joint process distribution
with respect to a new probability measure $\widehat P$ when they do not under $P$.
This facilitates efficient simulation as will become clear in the sequel.
(The reader can take $L^j=L=1$ on first reading so we are back to the
situation considered in Longstaff \& Schwartz 2001.)
Now, we generalize $A^N_t$ and $b^N_t$ to
\begin{eqnarray}
A^N_t=\frac1N \sum_{j=1}^N A_j, \text{ where } A_j=
\frac{L_t^je^J(S^j_t,V^j_t)e^J(S^j_t,V^j_t)' 1_{Z^j_t>0}}{\frac1N \sum_{i=1}^N1_{Z^i_t>0}},\\ 
b^N_t=\frac1N \sum_{j=1}^N b_j, \text{ where } b_j=\frac{L_t^j Z^j_{\tau^{J,j}_{t+1}}e^J(S^j_t,V^j_t) 1_{Z^j_t>0}}{\frac1N \sum_{i=1}^N1_{Z^i_t>0}}.
\end{eqnarray}
The standard i.i.d.\ strong law does not apply since
$\tau^{J,j}_{t+1}$ depends weakly on the other particles through the projection estimate.
Still, this dependence dies out fast enough as $N\rightarrow\infty$ that a general strong law does apply.
In particular, it follows from the (exchangeable) strong law of large numbers
that
\begin{description}
\item[\!\!\!\!SLLN-A]		
	\!$\displaystyle \lim_{N\rightarrow\infty}A^N_t\!=\!\frac{E[L_te^J(S_t,V_t)e^J(S_t,V_t)'1_{Z_t>0}]}{P(Z_t>0)}=\!\frac{\widehat E[e^J(S_t,V_t)e^J(S_t,V_t)'1_{Z_t>0}]}{P(Z_t>0)}$
\item[\!\!\!\!SLLN-b]
	\!$\displaystyle \lim_{N\rightarrow\infty}b^N_t\!=\!\frac{E[L_tZ_{\tau^{J}_{t+1}}e^J(S_t,V_t)1_{Z_t>0}]}{P(Z_t>0)}=\!\frac{\widehat E[Z_{\tau^{J}_{t+1}}e^J(S_t,V_t)1_{Z_t>0}]}{P(Z_t>0)}$,
\end{description}
where $\frac{d\widehat P}{dP}\bigg|_{\mathcal F_t}=L_t$ and
$\widehat E$ denotes expectation with respect to new probability measure $\widehat P$.
Under similar conditions \cite{Kouritzin:1996}
establishes that $\displaystyle \lim_{N\rightarrow\infty}\alpha^{J,N}_t=\alpha^J_t$ a.s. [$\widehat P$]
(and therefore a.s. [$P$])\ for any $\gamma>0$, where $\alpha^{J,j}_t$ is defined recursively by:
$\alpha^{J,0}_t=0$ and $k=1$ initially and then for $j=1,2,...,N$:
\begin{equation}
(\alpha^{J,j}_t,k)=\left\{\begin{array}{cc}(\alpha^{J,j-1}_t,k)&Z_t^j=0\\
(\alpha^{J,j-1}_t+\frac{\gamma L_t^j}k( Z^j_{\tau^{J,j}_{t+1}} -e^J(S^j_t,V^j_t)' \alpha^{J,j-1}_t)e^J(S^j_t,V^j_t),k+1)&Z_t^j>0\end{array}\right..
\end{equation}
Recall here that $(S,V,Z)$ has the desired distribution under $\widehat P$ not $P$ so
\begin{equation}
\alpha^J_t=\widehat E[e^J(S_t,V_t)e^J(S_t,V_t)'1_{Z_t>0}]^{-1}\widehat E[Z_{\tau^{J}_{t+1}}e^J(S_t,V_t)1_{Z_t>0}].
\end{equation}
(The triangle nature (through dependence of $\alpha^{J,j}_t$ on the number of particles $N$ of the summed terms in $b^N_t$) was not considered in this work.  However, the proof will still work in this case.)
Hence, we obtain convergence to the same solution as the least-squares
regression method but without numerically nasty matrix inversion.
Substituting $\displaystyle \lim_{N\rightarrow\infty}\alpha^{J,N}_t=\alpha^J_t$ a.s.\ into the work of \cite{ClPr} yields (after a small amount
of work) convergence in probability (at least) for this option pricing procedure.
Moreover, \cite{KoSa15} and \cite{Kouritzin:1994} could be used to obtain a.s.\ rates of convergence
and rates of $r^{th}$-mean convergence respectively for our parameter estimates if Conditions SLLN-A and SLLN-b are replaced by slightly stronger conditions (that would still hold in our setting).

Our first contribution is a numerically stable alternative to 
LSM algorithm of \cite{LoSc}. 
In particular, the following SA algorithm will be a big improvement when
$J$ is not very small.

\vspace*{0.3cm}
\par\noindent
{\bf Initialize:} Fix functions $e_k$ and $\gamma>0$; set $\zeta=\lambda=0$, all $\alpha^{J}_t=0$ and all $\tau^{J,j}=T$.\\ 
{\bf Simulate:} 
Create independent copies $\{L^j,S^j,V^j,Z^j\}_{j=1}^N$ of $(L,S,V,Z)$. \\
{\bf Repeat:} for $t=T-1$ down to $0$:\\
\indent $k=0$\\
\indent{\bf Repeat:} for $j=1$ to $N$:\\
\indent \indent {\bf Stochastic Approximation:}
If $Z_t^j>0$ then $k=k+1$ and
\begin{equation}
\alpha^{J}_t=\alpha^{J}_t+\frac{\gamma L_t^j}k( Z^j_{\tau^{J,j}} -e^J(S^j_t,V^j_t)' \alpha^{J}_t)e^J(S^j_t,V^j_t)
\end{equation}
\\
\indent{\bf Repeat:} for $j=1$ to $N$:\\
\indent \indent {\bf Adjust Stopping Times:}
If $Z_t^j>0$ and $Z_t^j\ge \alpha^{J}_t\cdot e^J(S^j_t,V^j_t)$, then $\tau^{J,j}=t$
\\
{\bf Price Option:}\\
\indent {\bf Repeat:} for $j=1$ to $N$:\\
\indent\indent $\zeta=\zeta+L^j_{\tau^{J,j}} Z^j_{\tau^{J,j}}$\\
\indent\indent $\lambda=\lambda+L^j_{\tau^{J,j}}$\\
\indent {\bf Value:} $O=\frac{\zeta}{\lambda}$\\

\begin{RM}
\emph{For each $\{(L^j_t,S^j_t,V^j_t,Z^j_t),\ t=0,1,...,T\}$,
$L^j$ is a non-negative martingale with mean $1$,
$\{(S_t^j,V_t^j),\ t=0,1,...,T\}$ has the desired risk-neutral (process) distribution and $\{Z_t^j,\ t=0,1,...,T\}$ is the 
discounted payoff process
with respect to probability $\widehat P^j(A)=E[L_T^j 1_A]$.
The preferred method to create these simulations for the Heston and other models with explicit
weak solutions follows in Subsection \ref{HestSimAlg}.
In this case, $L^j_t={\widehat L}^j_{t\wedge \eta_\varepsilon}$ where $\widehat L^j$ and $\eta_\varepsilon$
are defined in Subsection \ref{HestSimAlg}.
}
\end{RM}

\begin{RM}
\emph{This procedure is set up to be convenient for American options.
However, it is easy to adjust it to Asian options.
If this is desired, then we would simulate the running average price $R^j_t$ as
well (see Remark \ref{Thm12Use} to follow). 
These average prices would
become the $S^j$'s in this procedure, while the spot price would become
part of the $V^j$'s.
For example, in our Heston case each $V^j$ would be the whole $2$-dimensional
model and the \emph{new} $S^j$ would just be the average price as explained
in  Remark \ref{Thm12Use}.
}
\end{RM}

\begin{RM}
\emph{The SA algorithm gain $\gamma>0$ can effect performance due to
the finiteness of our particle system.
We choose a reasonable scalar $\gamma$.
However, a more general step size $\gamma/{k^\alpha}$ in place of $\gamma/{k}$
(see Kouritzin \& Sadeghi 2015 for a discusssion), a (positive definite) matrix-valued $\gamma$ or a two step algorithm
	like that introduced in \cite{PoJu92} may improve performance further.} 
\end{RM}

\begin{RM}
\emph{The first $J$ Haar bases functions on $[0,K]$ can be a good choice of $(e^S_k)_{k=1}^J$ for
a price only model and a put option with strike price $K$.
For volatility in Heston-type models, we can adapt the Haar bases to $[0,\infty]$.
Specifically, letting $h_k$ be the $k^{th}$ Haar function on $[0,1]$, we can rescale
by letting $e^V_k(x)=\sqrt{s'(x)}h_k\left(s(x)\right)$ for some differentiable \emph{scale} function
$s$ satisfying $s(0)=0$ and $\lim_{x\rightarrow\infty}s(x)=1$ to obtain new
bases functions $\{e^V_k\}_{k=1}^J$ on $[0,\infty]$.
An example is $s(x)=\frac{x}{1+x}$ so $e^V_k(x)=\frac1{1+x}h_k\left(\frac{x}{1+x}\right)$.
Naturally, there are other good scalings and choices of $(e^V_k)$.
Indeed, we will use the weighted Laguerre functions below since that is what \cite{LoSc} used.
	}
\end{RM}

\begin{RM}
\emph{We call this algorithm the SA or SA pricing algorithm.
Our version of the LSM algorithm is obtained simply by replacing the Stochastic Approximation part
by the following Least Squares Regression:\\
\indent $k=0$\\
\indent{\bf Repeat:} for $j=1$ to $N$:\\
\indent \indent {\bf Least Squares Regression:}
If $Z_t^j>0$ then $k=k+1$ and
\begin{eqnarray}
	A^{J}_t&=&\frac{k-1}kA^{J}_t+\frac{L_t^j}ke^J(S^j_t,V^j_t)e^J(S^j_t,V^j_t)'\\
	b^{J}_t&=&\frac{k-1}kb^{J}_t+\frac{L_t^j}k Z^j_{\tau^{J,j}} e^J(S^j_t,V^j_t)
\end{eqnarray}
\indent $\alpha^{J}_t=(A^{J}_t)^{-1}b^{J}_t$.\\
We also set all $A^{J}_t=0$ (matrix of all zeros) and $b^{J}_t=0$ during the initialization.
The rest of the algorithm is the same.
	}
\end{RM}

\subsection{Explicit and Weighted Solutions}

There are several papers on exact simulation for the Heston model (see e.g.\ Andersen 2007, van Haastrecht \& Pelsser 2010).
Most of these contributions build off of \citet{BroadieKaya:2006} 
and/or rely on a change of variables as well as Feller's 
characterization of the transition function for the square root diffusion.
Generic difficulties of these methods are:  
(a) algorithm complexity - often involving numeric convergence,		
(b) accommodating all possibly desired drifts,
(c) allowing derivative payoffs that depend on the underlying asset at many points in time,
(d) admitting time dependence in the spot price variance,	
and (e) handling the volatility approaching or hitting $0$. 

Alternatively, one should consider the possibility of  
\emph{explicit} representations of the Heston SDEs as a time-dependent 
function $\phi\left(\int_s^t U_u dW_u,t\right)$ of a simple Gaussian
stochastic integral.
It is discovered in Theorem 1 of our companion paper \citet{Kouritzin/Remillard:2015} 
that a necessary and sufficient condition for the SDE
\begin{equation}
dX_{t}=b(X_{t})dt+\sigma (X_{t})dW_{t},  \label{sde1}
\end{equation} 
to have
a strong solution with such an explicit representation locally (for some drift coefficient
$b$) is the diffusion coefficient columns $\sigma_j$ satisfy the Lie bracket condition: 
\begin{equation}\label{Bracket1} 
(\nabla \sigma_i)\sigma_j=(\nabla \sigma_j)\sigma_i\ \ \forall i,j.
\end{equation} 
(This theorem from Kouritzin \& Remillard 2016 was motivated
in part by the works of Doss 1977, Sussmann 1978, Yamato 1979, Kunita 1984, Kouritzin
\& Li 2000 and Kouritzin 2000
that also express SDE solutions in terms of the driving Brownian motion.)
Unfortunately, the Heston model does not satisfy (\ref{Bracket1}) since
\begin{eqnarray}
(\nabla \sigma_1)\sigma_2=\left(\!\begin{array}{c}sv\rho\sqrt{1-\rho^2}+\frac{s\kappa\sqrt{1-\rho^2}}{2}\\0\end{array}\!\right)\ne\left(\!\begin{array}{c}sv\rho\sqrt{1-\rho^2}\\0\end{array}\!\right) =(\nabla \sigma_2)\sigma_1
\end{eqnarray}
when 
\begin{equation}
\sigma=\left(\sigma_1\,\sigma_2\right)=\left(\begin{array}{cc}\sqrt{1-\rho^2}sv^{\frac{1}{2}}&\rho sv^{\frac{1}{2}}\\0&\kappa v^{\frac{1}{2}}\end{array}\right),
\end{equation}
where $s$ and $v$ represent the state variables for price and variance (square of volatility).
Hence, we will have to consider weak solutions to get an explicit representation
for the Heston SDEs.
While our focus herein is largely on solving the SDEs and using the solutions
in simulation for option pricing, the solutions can also be used in other ways.

Explicit solutions are fragile.
For example, it is shown in \citet{Kouritzin:2000} that scalar SDEs
only have explicit solutions for specific drift coefficients.
Hence, it is reasonable to expect a condition on the
Heston model parameters for an explicit solution (if they
are even possible).
This condition is:
\begin{description}
\item[C] $\nu=\frac{n\kappa^2}4$ for some $n=1,2,3,...$.
\end{description}
Fortunately, this is all that is needed.

\begin{thm} \label{Theorem1}Suppose $n\in\{1,2,3,4,...\}$, Condition (C) holds with this
$n$ and $W^1,...,W^n,B$ are independent standard Brownian motions.
Then, the Heston (price and volatility) model (\ref{Heston}) has explicit weak solution: 
\begin{eqnarray}\label{ExplicitSt}
\!\!\!\!\!\!\!\!\!\!\!\!\!S_t&\!\!\!\!=&\!\!\!\! S_0
\exp\!\bigg(\!\sqrt{1\!-\!\rho^2}\!\int_0^t \!V_s^\frac12 dB_s\!
+\!\left[\mu\!-\frac{\nu\rho}\kappa\right]\! t\!
+\!\left[\frac{\rho\varrho}\kappa-\frac12\right]
\!\int_0^t \!V_s ds+\frac\rho\kappa (V_t-\!V_0)\!
\bigg),
\\\label{ExplicitVt} \!\!\!\!\!\!\!\!\!V_t&\!\!\!\!=&\!\!\!\! \sum_{i=1}^n(Y_t^i)^2,
\end{eqnarray}
where $\{Y_t^i=\frac{\kappa}2\int_0^t e^{-\frac{\varrho}2 (t-u)}dW^i_u+e^{-\frac{\varrho}2t}Y^i_0\}_{i=1}^n$
are Ornstein-Uhlenbeck processes and
\begin{equation}
	\beta_t=\sum_{i=1}^n \int_0^t\frac{Y_u^i}{\sqrt{\sum_{j=1}^n(Y_u^j)^2}}dW^i_u
\end{equation}
is the other Brownian motion appearing in (\ref{Heston}).
\end{thm}
While the drift and diffusion coefficients do not satisfy the classical conditions for a strong
solution, it follows from Remark 1.1 of \cite{BaPe} as well as \cite{RoWi:1987} that it does
have a weak solution.
Theorem \ref{Theorem1} also establishes weak solutions but, most importantly,
also gives them explicitly in a computable way.
\proof See Appendix. \qed
\begin{RM}
\emph{The solution is valid for \emph{any} $\{Y_0^i\}_{i=1}^n$ such that $\sum_{i=1}^n(Y_0^i)^2= V_0$.
By expanding the squares, $V_t$ can be written as $V_t=V^\chi_t+V_t^G+V_t^D$,
the sum of a $\chi^2$ random variable plus a Gaussian variable plus a deterministic piece.
In particular, the moment generating functions of the first two pieces are:
\begin{equation}\label{Vrandoms}
M_{V^\chi_t}(\theta)=\left(\!1-\frac{\kappa^2}{2\varrho}\left[1-e^{-\varrho t}\right]\theta\!\right)^{-\frac{n}2}
\text{and }
M_{V^G_t}(\theta)=\exp\left(\frac{V_0\kappa^2}{2\varrho e^{\varrho t}}\left[1-e^{-\varrho t}\right]\!\theta^2\!\right)
\end{equation}
(for $\theta$ in a neighbourhood of $0$) while the deterministic piece is just 
\begin{equation}\label{Vdeterm}
V^D_t=\exp(-\varrho t)V_0
\text{.}
\end{equation}
Then, it follows by the Burkholder-Davis-Gundy inequality, Jensen's inequality, Fubini's theorem
and the moment bounds for the $\chi^2$ and Gaussian random variables that
there is a $C_{r,t}>0$ such that
\begin{equation}
E\left[\left|\int_0^t V_s^\frac12 dB_s\right|^r\right]
\le C_{r,t} E\left[\int_0^t\left| V_s \right|^\frac{r}2 ds\right]<\infty
\end{equation}
for any $r\ge 2$, $t>0$ and $\int_0^t \!V_s^\frac12 dB_s$ is an $L^r$-martingale for any $r>0$.
}
\end{RM}
\begin{RM}
\emph{One can apply It\^{o}'s formula to (\ref{ExplicitSt}) and (\ref{ExplicitVt})	
to verify they do indeed satisfy (\ref{Heston}).
Hence, one could have just guessed this solution and then
checked it.
However, nobody every has and it took the development in the appendix for the author to formulate this solution.
}
\end{RM}
Noting that mathematical models are just approximations of reality, one
can sometimes justify picking a Heston model such that Condition (C) is true.
We demonstrate simulation for this case in the next section. 
However, we also want a solution for other parameters not just those satisfying Condition (C).
With this in mind, we first define the \emph{Closest Explicit Heston} case:
\begin{eqnarray}\label{SimpHeston}
d\left(\begin{array}{c}\widehat S_{t}\\\widehat V_{t}\end{array}\right)
=\left(\begin{array}{c}\mu_\kappa \widehat S_{t}\\\nu_\kappa-\varrho \widehat V_{t}\end{array}\right)dt+\left(\begin{array}{cc}\sqrt{1-\rho^2}\widehat S_{t}\widehat V_{t}^{\frac{1}{2}}&\rho \widehat S_{t}\widehat V_{t}^{\frac{1}{2}}\\0&\kappa \widehat V_{t}^{\frac{1}{2}}\end{array}\right)\left(\begin{array}{c}dB_{t}\\d\widehat \beta_{t}\end{array}\right),\\ 
\label{numuk}\text{where }n=\left\lfloor\frac{4\nu}{\kappa^2}+\frac12\right\rfloor\vee 1,\ \nu_\kappa=\frac{n\kappa^2}4,\ 
\mu_\kappa=\mu+\frac{\rho}{\kappa}\left(\nu_\kappa-\nu\right),
\end{eqnarray}
where Condition (C) is valid (with $\nu=\nu_\kappa$).
Then, we re-weight the outcomes of the closest explicit Heston to get general Heston solutions.
\begin{RM}
\emph{Finding the closest explicit Heston solution amounts to
selecting $n$.}
\end{RM}
The general Heston model (\ref{Heston}) without Condition (C) also has an explicit weak solution
with respect to some new probability until the volatility drops too low.
\begin{thm} \label{Theorem2}
Let $\varepsilon\in(0,1)$, $T>0$, $(\Omega,\mathcal F,\{\mathcal F\}_{t\in [0,T]},P)$ be a filtered probability space, $V_0, S_0$ be given random variables with $V_0>\varepsilon$,
$\{W^1,...,W^n,B\}$ be independent standard Brownian motions with respect to $(\Omega,\mathcal F,\{\mathcal F\}_{t\in [0,T]},P)$,
\begin{eqnarray}\label{ExplicitStful}
\!\!\!\!\!\!\!\!\!\!\!\!\widehat S_t&\!\!\!\!=&\!\!\!\! S_0
\exp\!\bigg(\!\sqrt{1\!-\!\rho^2}\!\int_0^t \!\widehat V_s^\frac12 dB_s\!
+\!\left[\mu\!-\frac{\nu\rho}\kappa\right]\! t\!
+\!\left[\frac{\rho\varrho}\kappa-\frac12\right]
\!\int_0^t \!\widehat V_s ds+\frac\rho\kappa (\widehat V_t-\!\widehat V_0)\!\!
\bigg)\ \ 
\\\label{ExplicitVtful} \!\!\!\!\!\!\!\!\widehat V_t&\!\!\!\!=&\!\!\!\! \sum_{i=1}^n(Y_t^i)^2,
\ \ \eta_\varepsilon=\inf\left\{t:\widehat V_t\le\varepsilon\right\}\ \text{ and}\\
\!\!\!\!\!\!\!\!\label{L2way}
\widehat L_t&\!\!\!=&\!\!\!\!\exp\left\{ \frac{\nu-\nu_\kappa}{\kappa^2} \left[\ln(\widehat V_t)-\ln(\widehat V_0)
+\int_0^t\frac{\kappa^2-\nu_\kappa-\nu}{2\widehat V_s}+\varrho\, ds
\right] \right\},
\end{eqnarray}
where $Y_t^i=\frac{\kappa}2\int_0^t e^{-\frac{\varrho}2 (t-u)}dW^i_u+e^{-\frac{\varrho}2t}Y^i_0$ for $i=1,2,...,n$.
Define 
\begin{eqnarray}
\label{Bbetadef}
\!\!\!\!\!\!\!\beta_t&\!\!=&\!\!\sum_{i=1}^n \int_0^t\frac{Y_u^i}{\sqrt{\sum_{j=1}^n(Y_u^j)^2}}dW^i_u+\int_0^{t\wedge\eta_\varepsilon} \frac{\nu-\nu_\kappa}{\kappa \widehat V_s^\frac12}ds,\ \ \text{and}\\
\!\!\!\!\!\!\!\widehat P(A)&\!\!=&\!\!E[1_A\widehat  L_{T\wedge\eta_\varepsilon}]\ \ \forall A\in \mathcal F_T.
\end{eqnarray}
Then, $\eta_\varepsilon$ is a stopping time and $\widehat L_{t\wedge\eta_\varepsilon}$ is a
$L^r$-martingale with respect to $P$ for any $r>0$.
Moreover, $(B,\, \beta)$ are independent standard Brownian motions and
\begin{eqnarray}\label{WeakHestonPtilde}
d\!\left(\!\begin{array}{c}\widehat S_{t}\\\widehat V_{t}\end{array}\!\right)
=
\left\{\!\!\begin{array}{cc}
\left(\!\begin{array}{c}\mu\widehat S_{t}\\\nu-\varrho \widehat V_{t}\end{array}\!\right)dt+\left(\!\begin{array}{cc}\sqrt{1-\rho^2}\widehat S_{t}\widehat V_{t}^{\frac{1}{2}}&\rho \widehat S_{t}\widehat V_{t}^{\frac{1}{2}}\\0&\kappa \widehat V_{t}^{\frac{1}{2}}\end{array}\!\right)\left(\!\begin{array}{c}dB_{t}\\d \beta_{t}\end{array}\!\right),
& t\le \eta_\varepsilon\\
\left(\!\begin{array}{c}\mu_\kappa \widehat S_{t}\\\nu_\kappa-\varrho \widehat V_{t}\end{array}\!\right)dt+\left(\!\begin{array}{cc}\sqrt{1-\rho^2}\widehat S_{t}\widehat V_{t}^{\frac{1}{2}}&\rho \widehat S_{t}\widehat V_{t}^{\frac{1}{2}}\\0&\kappa \widehat V_{t}^{\frac{1}{2}}\end{array}\!\right)\left(\!\begin{array}{c}dB_{t}\\d \beta_{t}\end{array}\!\right),
& t> \eta_\varepsilon
\end{array}\right.		
\end{eqnarray}
on $[0,T]$ with respect to $\widehat P$.
\end{thm} 
\proof See Appendix. \qed
\par\noindent{\bf Notation:} We are using $\widehat S,\widehat V$ for solutions to the \emph{closest explicit} Heston model, reserving
$S,V$ for the general case. 
Henceforth, we will use 
\begin{equation}
\widehat \beta_t=\sum\limits_{i=1}^n \int_0^t\frac{Y_u^i}{\sqrt{\sum_{j=1}^n(Y_u^j)^2}}dW^i_u\ \mbox{ and } \beta_t=\widehat \beta_t+\int_0^{t\wedge\eta_\varepsilon} \frac{\nu-\nu_\kappa}{\kappa \widehat V_s^\frac12}ds.
\end{equation}
\begin{RM}
\emph{	
With respect to the manufactured measure $\widehat P$, 
$(\widehat S_t,\widehat V_t)$ satisfies the general Heston model (\ref{Heston}) 
until the stopping time $\eta_\varepsilon$ and then the Closest Explicit Heston model
(\ref{SimpHeston}) after that.	
Conversely, since $\mu-{\nu\rho}/\kappa=\mu_\kappa\!-{\nu_\kappa\rho}/\kappa$,
we find that
$(\widehat S,\widehat V)$ satisfies (\ref{SimpHeston}) for all $t\in[0,T]$ with respect to $P$ 
by (\ref{ExplicitStful}) and Theorem \ref{Theorem1}.
}
\end{RM}
Our first concern about Theorem \ref{Theorem2} is:
The desired solution here is only good until $\eta_\varepsilon$, i.e.\ until the volatility drops
too low (or we hit the final `simulation time' $T$).
From a finance viewpoint, one can ask: ``Is it realistic that the volatility
of my asset drops to zero any way?''.
Usually, this constraint of not being able to simulate through essentially deterministic price
change is not a practical issue and, even if this happens, we just fall back to the
closest explicit alternative.
Our second concern is:
The desired solution here is with respect to
a manufactured probability $\widehat P$.
However, 
\begin{enumerate}
\item
this manufactured-probability solution is ideal for option pricing calculations,
\item
this manufactured-probability solution is also excellent for pricing derivatives via Monte-Carlo-type simulation.
\end{enumerate}
To illustrate the last point, we suppose we have independent copies
$\{(\widehat S^j,\widehat V^j,\widehat L^j)\}_{j=1}^N$ of $(\widehat S,\widehat V,\widehat L)$ and
$\eta^j_\varepsilon=\inf\left\{t:\widehat V^j_t\le\varepsilon\right\}$. 
Then, using the law of large numbers (for weakly-dependent variables) and $\widehat L$'s martingale property
\begin{eqnarray}\label{WSLLN}
\frac1N\sum_{j=1}^N \widehat L^j_{t\wedge\eta_\varepsilon^j}g(\widehat S^j_{[0,t]},\widehat V^j_{[0,t]},\Delta^N_{[0,t]})	
&\!\!\!\rightarrow &\!\!\!E[\widehat L_{t\wedge\eta_\varepsilon}g(\widehat S_{[0,t]},\widehat V_{[0,t]},\Delta_{[0,t]})]
\\\nonumber
&\!\!\!=&\!\!\!\widehat E[g(\widehat S_{[0,t]},\widehat  V_{[0,t]},\Delta_{[0,t]})]
\end{eqnarray}
for any bounded, measurable function $g$ and $t\le T$,
where $\widehat E$ denotes expectation with respect to $\widehat P$,
$\Delta^N_{[0,t]}$ is the empirical process $\frac1N\sum_{j=1}^N \delta_{\widehat L^j_{[0,t]},\widehat S^j_{[0,t]},\widehat V^j_{[0,t]}}$ and
$\Delta_{[0,t]}$ is the joint distribution of
$(\widehat L_{[0,t]},\widehat S_{[0,t]},\widehat V_{[0,t]})$.
(Here, $\widehat L_{[0,t]},\widehat S_{[0,t]},\widehat  V_{[0,t]}$ denote the paths of $\widehat L,\widehat  S,\widehat  V$ over
$[0,T]$ held constant after $t$.)
(\ref{WSLLN}) is what we need for (SLLN-A, SLLN-b) and therefore to use
$\{(\widehat S^j,\widehat V^j,\widehat L^j)\}_{j=1}^N$ in our SA Pricing Algorithm of 
the previous subsection.
In the next subsection, we reduce these theorems to useful algorithms
that can be used for simulation or within the LSM and SA option-pricing algorithms.
\begin{ex}
\emph{	
For pricing an American call option with strike price $K$, we would use	
$g(\widehat S^j_{[0,t]},\widehat V^j_{[0,t]},\Delta^N_{[0,t]})=e^{-\mu \tau^{J,j}_0}(\widehat S_{\tau^{J,j}_0}^j-K)\vee0$,
where $\tau^{J,j}_0$ satisfies (\ref{tauJjdef}) in the LSM algorithm or a
similar formula (with slightly different but still asymptotically consistent
coefficients $\alpha^{J,N}_t$) in the SA algorithm.
Since $\tau^{J,j}_0$ depends upon the paths of $\widehat S$ and $\widehat V$ so does 
$g(\widehat S^j_{[0,t]},\widehat V^j_{[0,t]},\Delta^N_{[0,t]})$ in American (and Asian) option pricing
examples.
Since $\tau^{J,j}_0$ uses projection estimates that depend on the other particles, we
have to include the empirical process $\Delta^N_{[0,t]}$, which results in
weakly interacting variables instead of independent ones.
To justify the weakly-interacting SLLN in this example, we note from previous discussion that
the projection estimates converge to the desired projection, which no longer depends upon
the other particles.
Also, the exact dependence of $\tau^{J,j}_0$ on the other particles and the paths is
not critical but can be determined from the SA algorithm and the Weighted Heston Algorithm to
follow.
}
\end{ex}

\subsection{Weighted and Explicit Heston Simulation}\label{HestSimAlg}

Defining constants 
\begin{equation}
a=\sqrt{1-\rho^2},\,b=\mu-\frac{\nu\rho}{\kappa},\,c=\frac{\rho\varrho}{\kappa}-\frac{1}{2},\,
d= \frac{\rho}{\kappa},\,e=\frac{\nu-\nu_\kappa}{\kappa^2},\,f=e\frac{\kappa^2-\nu-\nu_\kappa}{2}, 
\end{equation}
we find that (\ref{ExplicitStful},\ref{L2way}) can be rewritten as
\begin{eqnarray}\label{Stfulabcd}
\!\!\!\!\!\!\!\!\!\!\!\!\widehat S_t&\!\!\!\!=&\!\!\!\! \widehat S_{t-1}
\exp\!\bigg(\!a\int_{t-1}^t \!\widehat V_s^\frac12 dB_s\!
+b +c \int_{t-1}^t \!\widehat V_s ds+d\ (\widehat V_t-\!\widehat V_{t-1})\!\bigg)
\\
\!\!\!\!\!\!\!\!\label{Lef}
\widehat L_t&\!\!\!=&\!\!\!\!\widehat L_{t-1}\exp\left\{ e \left(\ln\left(\frac{\widehat V_t}{\widehat V_{t-1}}\right)+\varrho\right) 
+f\int_{t-1}^t\frac{1}{\widehat V_s}\,ds
 \right\}.
\end{eqnarray}
The stochastic integral in (\ref{Stfulabcd}) is conditionally (given $\widehat V$) Gaussian since $\widehat V$ 
and $B$ are independent so simulation is just a centered normal random variable with
variance $a^2\int_{t-1}^t \!\widehat V_s ds$.
Even the weight (\ref{Lef}) avoids stochastic integrals.
There are a number of choices for the two deterministic integrals 
to be computed like:
\begin{itemize}
\item[T:] 
$\displaystyle \int_{t-1}^t \!\widehat V_s ds\approx \frac1{2M}\left\{\widehat V_{t-1}+\widehat V_t+2\sum_{l=1}^{M-1}\widehat V_{t-\frac{l}M}\right\}$
\item[S $\frac13$:]		
$\displaystyle \int_{t-1}^t \!\widehat V_s ds\approx \frac1{3M}\left\{\widehat V_{t-1}+\widehat V_t+2\sum_{l=1}^{\frac{M}2-1}\widehat V_{t-\frac{2l}M}+4\sum_{l=1}^{\frac{M}2}\widehat V_{t-\frac{2l-1}M}\right\}$
\item[S $\frac38$:]		
$\displaystyle \int_{t-1}^t \!\widehat V_s ds\approx \frac3{8M}\left\{\widehat V_{t-1}+\widehat V_t+2\sum_{l=1}^{\frac{M}3-1}\widehat V_{t-\frac{3l}M}+3\sum_{l=1}^{\frac{M}3}\widehat V_{t-\frac{3l-2}M}+3\sum_{l=1}^{\frac{M}3}\widehat V_{t-\frac{3l-1}M}\right\}$
\end{itemize}
(for the Trapezoidal, Simpson's $1/3$ and Simpson's $3/8$ rules respectively).
Similar formulae are also used for $\int_{t-1}^t \!1/{\widehat V_s} ds$.
Naturally, all of these will converge to the integral as $M\rightarrow\infty$.
$V$ does not satisfy the necessary smoothness conditions for the classical
errors of these numeric integral methods so it is unknown which will perform better.
Indeed, simulations will show there is very little difference on our examples.
Finally, it will be notationally convenient to restrict to the case $n$ is even
(the odd case is a minor modification) and to define three more constants
\begin{equation}
\sigma=\kappa\sqrt{\frac{1-e^{-\frac\varrho2}}{4\varrho}},\ \alpha=e^{-\frac\varrho4} \text{ and }n_2=\frac{n}2.
\end{equation}
The algorithm (with the hats removed for notational ease) is now as follows:
\vspace*{0.3cm}
\par\noindent
{\bf Initialize:} $\left\{(S_0^j,L_0^j,\eta^j_\varepsilon)=(S_0,1,T)\right\}_{j=1}^N$, $\left\{Y_{0}^{j,i}=\sqrt{\frac{V_{0}}{n}}\right\}_{j,i=1}^{N,n}$.
\vspace*{0.3cm}
\par\noindent
{\bf Repeat:} for times $t=1,2,...,T$ do\vspace*{0.3cm}\\
	\vspace*{0.3cm}
\indent {\bf Repeat:} for particles $j=1,2,...,N$ do\\
	\vspace*{0.3cm}
	\indent (1) $V_{t-\frac12}^j=0$, $V_t^j=0$\\ 
	\indent\ (2) {\bf Repeat:} for $i=1,2,...,n_2$ do\vspace*{0.3cm}\\ 
\indent \indent (a) Draw $[0,1]$-uniform $U_1,U_2,U_3,U_4$\\	
\indent \indent (b) $Y^{j,2i-1}_{t-\frac12}=\alpha Y^{j,2i-1}_{t-1}+\sigma\sqrt{-2\log U_{1}}\cos (2\pi U_{2})$ (Use Box-Meuller for normals)\\
\indent \indent (c) $Y^{j,2i}_{t-\frac12}=\alpha Y^{j,2i}_{t-1}+\sigma\sqrt{-2\log U_{1}}\sin (2\pi U_{2})$\\
\indent \indent (d) $Y^{j,2i-1}_t=\alpha Y^{j,2i-1}_{t-\frac12}+\sigma\sqrt{-2\log U_{3}}\cos (2\pi U_{4})$\\
\indent \indent (e) $Y^{j,2i}_t=\alpha Y^{j,2i}_{t-\frac12}+\sigma\sqrt{-2\log U_{3}}\sin (2\pi U_{4})$\\
\indent \indent (f) $V_{t-\frac12}^j=V_{t-\frac12}^j+(Y^{j,2i-1}_{t-\frac12})^2+(Y^{j,2i}_{t-\frac12})^2$, $V_t^j=V_t^j+(Y^{j,2i-1}_t)^2+(Y^{j,2i}_t)^2$\vspace*{0.3cm}\\
\indent\ (3) Set $IntV^j=\frac{V^j_{t-1}+4V^j_{t-\frac12}+V^j_t}6$ (Simpson's $\frac13$ rule, $M=2$)\\	
\indent\ (4) Set $N^j=\mathcal N\left(0,a\sqrt{ IntV^j}\right)$ (centered normal RV)\\	
\indent\ (5) $S_t^j=S_{t-1}^j\exp(N^j+b+c\,IntV^j +d\ (V^j_t-V^j_{t-1}))$\\ 
\indent\ (6) $Z_t^j=p(t,S_t^j)$\ (Discounted Payoff e.g. $e^{-\mu t}(K-S_t^j)\vee0$ for American put)\\ 
\indent\ (7) If $t\le \eta^j_\varepsilon$ then\\
\indent \indent \ If $V^j_{t-\frac12}\wedge V^j_t>\varepsilon$ then $L_t^j=L_{t-1}^j\exp\left\{ e \left(\ln\left(\frac{V^j_t}{V^j_{t-1}}\right)+\varrho \right)
+\frac{f}6\left[\frac{1}{V^j_{t-1}}+\frac{4}{V^j_{t-\frac12}}+\frac{1}{V^j_t}\right]
 \right\}$\\ 
\indent \indent \ Otherwise $\eta^j_\varepsilon=t-1$\\ 

\begin{RM}\label{Thm12Use}
\emph{
There are some practical notes about using this algorithm:
\begin{enumerate}
\item $e^{-\mu}$ is the discount factor in (6) so $e^{\mu t}$ dollars at time $t$
are considered as valuable as $\$1$ at time $0$.
\item To price Asian options, where our payoff is in terms of the
running average price not the spot price, on the Heston model we initiate $R_0=0$, add a step:\\ 
(5a) $R_t^j=\frac{t-1}t R_{t-1}^j +\frac1t S_t^j$\\
and change the payoff process in (6) to $Z_t^j=p(t,R_t^j)$. 
You can then impose a ``lockout period'' by resetting the $Z_t^j$ to $0$ for those times.
\item	
In the Theorem 1 case of $\nu={n\kappa^2}/4$, we have explicit solutions without the need of weights.
In this case, we can skip Step (7) and remove all references to $\eta_\varepsilon$ and $L^j$ in this algorithm.
We call this reduced algorithm for Theorem 1 the \emph{Explicit Heston Simulation} algorithm
and the general algorithm (as stated above) for Theorem 2 the \emph{Weighted Heston Simulation}
algorithm.
\item
For added efficiency, Box-Meuller could be used in Step (4) as well.
Moreover, you could lump constants together to reduce multiplications (at
the cost of code readability).
We do not employ these added efficiencies herein.
\item
A larger $M$ or a better integral approximation could also be used to improve
performance in Step (3).
We used $M=2$ and Simpson's $1/3$ rule for algorithm clarity reasons only.
\end{enumerate}
}
\end{RM}

To understand the need to stop (at $\eta_\varepsilon$) before the volatility gets too small, 
we consider the situation where the volatility $V_t^\frac12=0$.
Then, the (closest explicit and general) Heston volatility equations become deterministic
\[
	d\widehat V_t=\nu_\kappa dt,\ \ \ \ dV_t=\nu dt
\]
and it is obvious which solution one has.
This makes model distributions singular to each other when $\nu_k\ne \nu$.


\section{Performance of Explicit Solution Simulation} \label{Sim-Comparison}

We compare our algorithms numerically to some of the more
popular methods, first in this section on simulation and then in the
next section on progressively 
more involved option pricing problems.
All experiments in both sections are performed on the same computer system, consisting of a
Lenovo X240s Laptop with a 4th generation Intel Core i7-4500U @ 1.80GHz processor,
8GB DIMML memory,
1TB 5400 RPM hard disk, Windows 8.1 64 bit operating system and the C++ compiler from Visual Studio professional 2013.

\subsection{Non-failure of Explicit Heston Simulation}

We will call a simulation where a \emph{negative} volatility is produced a \emph{failure} and
the first time this occurs is defined as the \emph{break time} $\tau$.
The Euler and Milstein methods both fail by producing negative volatility
values that can not be square rooted without change (like setting to zero). 
Conversely, our Explicit Heston algorithm can not fail in this manner
as the volatility is exact and stays non-negative by its construction.

First, suppose $\mu=0.0319,\rho=-0.7,\varrho=6.21,\kappa=0.61 $ and $\nu={\kappa^2}/4$
so the (SDE model) volatility can hit zero but can not go negative.
Our initial state is $S_0=100,V_0=0.010201$ and we run the simulation
either $10,000$ or $40,000$ times until final time $T=50$.
We use either $100$ or $200$ discretization steps between each integer
time.
The relative breaking frequency of Euler and Milstein simulations are shown in Tables \ref{table:Relative breaking frequency for n=1} below.

\begin{table}[H]
	\centering
	\begin{tabular}{|l|c|c|c|c|}
		\hline \bf Scheme& \multicolumn{2}{|c|}{Euler} &\multicolumn{2}{|c|}{Milstein}\\
		\hline 
		\bf N& $10,000$ & $40,000$ & $10,000$ & $40,000$\\
		\hline 
		\bf Steps & $100$ & $200$ & $100$ &$200$ \\ 
		\hline \bf $\tau\in(0,1]$ & 0.972386 & 0.972184 &  0.932158 &  0.914071 \\ 
		\hline \bf $\tau\in(1,2]$ & 0.026434 & 0.025734 & 0.062245 & 0.077341 \\
		\hline \bf $\tau\in(2,3]$ & 0.001134 & 0.001033 &  0.005166 &  0.007731 \\ 
		\hline \bf $\tau\in(3,4]$ & 0.000045 & 0.0000465 &  0.000394 &  0.000777 \\ 
		\hline \bf $\tau\in(4,5]$ & 0.000010 & 0.0000025 &  0.000037 &  0.0000713 \\ 
		\hline \bf $\tau>50$ & 0 & 0 &  0 &  0 \\ 
		\hline
	\end{tabular} 
	\caption{Relative breaking frequency for $\nu={\kappa^2}/{4}, \kappa=0.61, \varrho=6.21$.  Each column is a probability mass function (with some rows missing).  The top left value of 0.972386 indicates that a negative volatility
is expected 9,724 times within the first 100 steps until time 1 if 10,000 simulations are run many times from different seeds.
Similarly, the next value of 0.026434 indicates 264 of the 10,000 simulation
would survive until time 1 and then break in the 100 steps between times 1 and 2 on average.}
	\label{table:Relative breaking frequency for n=1}
\end{table}	
Ideally, there should not be any failures, so every simulation should exceed $\tau=T=50$ but actually none do.
One might think that this only happens when the volatility is supposed to hit zero.
However, increasing $\nu$ to ${\kappa^2}/2$, which is the critical or first case that the volatility should not hit 0, we still encounter the same problem, especially for the Euler scheme.
\begin{table}[H]
	\centering
	\begin{tabular}{|l|c|c|c|c|}
		\hline \bf Scheme& \multicolumn{2}{|c|}{Euler} &\multicolumn{2}{|c|}{Milstein}\\
		\hline 
		\bf N& $10,000$ & $40,000$ & $10,000$ & $40,000$\\
		\hline 
		\bf Steps & $100$ & $200$ & $100$ &$200$ \\ 
		\hline \bf $\tau\in(0,1]$ & 0.802964 & 0.767827 &  0.000492 &  0 \\ 
		\hline \bf $\tau\in(1,2]$ & 0.147584 & 0.165 & 0.000488 & 0 \\
		\hline \bf $\tau\in(2,3]$ & 0.037084 & 0.0.047847 &  0.000506 &  0 \\ 
		\hline \bf $\tau\in(3,4]$ & 0.009277 & 0.013768 &  0.000524 &  0 \\ 
		\hline \bf $\tau\in(4,5]$ & 0.002313 & 0.003941 &  0.000484 &  0 \\ 
		\hline \bf $\tau>50$ & 0 & 0 &  0.976822 &  1 \\ 
		\hline
	\end{tabular} 
	\caption{Relative breaking frequency for $\nu={\kappa^2}/{2}, \kappa=0.61, \varrho=6.21$. Each column is a probability mass function indicating empirical probability of breaking in each minute interval of either 100 or 200
	steps.}
	\label{table:Relative breaking frequency for n=2}
\end{table}	
For $\nu={\kappa^2}/2$, we see that Milstein scheme with 200 steps works well while the Euler scheme volatility still goes negative in every simulation.

\subsection{Comparison of Explicit Heston Simulation}

We provide an example of our Explicit Heston simulation 
and compare this to the traditional Euler and Milstein 
simulation methods.
In this approach, we create a \emph{ground truth} to judge performance from by fixing
Brownian paths $B,\beta$ and running the 
Milstein method once with the ridiculously small time step $\Delta t=1/2,000$.
We then used these fixed $B,\beta$ paths to calculate the error in the simulations 
discusssed in this subsection.
To get \emph{time} estimates we resort back to the normal efficient algorithms that would
be used in practice.
In this manner, we obtain comparable \emph{path-by-path} simulation error
with execution time estimates for the typical time it would take to produce
those errors.

For this example, we used the following collection of parameters: $\nu=\nu_\kappa={\kappa^2}/4,\mu=0.0319,\rho=-0.7,\varrho=6.21,\kappa=0.61 $ and $T=10$.
We also take the (non-ground-truth) Euler and Milstein time steps to be $\Delta t=1/M$,
where the number of steps are $M=200,\ 400,\ 1,000$.
Since Condition (C) holds we can remove all reference to $L$ and $\eta$
from the previously-given Heston simulation algorithm.
Tables \ref{table:Euler and Milstein} and \ref{table:Explicit solution} below show the performance and execution time of our Explicit
Heston algorithm with the Trapezoidal, Simpson's $1/3$ as well as Simpson's $3/8$ rule along with the Euler 
and Milstein methods.
For clarity, the performance is defined in terms of RMS error.
The RMS error for the Milstein method is:
\begin{equation}
	E^M=\sqrt{\frac 1N\sum_{t=1}^T\sum_{i=1}^N\left[(S_t^{M,i}-S_t^i)^2+(V_t^{M,i}-V_t^i)^2\right]},
\end{equation}
with $S^M,V^M$ being the price and volatility using the Milstein method and 
$S,V$ being the ground truth price and volatility.
The other RMS errors are defined similarly.
\begin{table}[H]
	\centering
	\begin{tabular}{|c|c|c|c|c|c|c|}
		\hline {} &\multicolumn{3}{|c|}{\bf Euler Scheme}  &\multicolumn{3}{|c|}{\bf Milstein Scheme} \\
		\hline \bf Steps & $200$ & $400$ & $1,000$ & $200$ & $400$ & $1,000$\\
		\hline \bf RMS & $18.8256$ & $14.1382$ & $9.79565$ & $10.5435$ & $7.08773$ & $4.2306$\\
		\hline \bf Time & $0.81$ & $1.672$ & $4.026$ & $0.936$ & $1.733$ & $4.731$\\
		\hline
	\end{tabular} 
	\caption{Comparison of RMS error from ground truth and computer execution time (in seconds).  The steps indicates the number of discretization steps per second.  The RMS error is averaged over all steps and for all 10 seconds as indicated in the formula above.  The numbers are also averaged over 20,000 seeds.}
	\label{table:Euler and Milstein}
\end{table}

\begin{table}[H]
	\centering
	\begin{tabular}{|c|c|c|c|c|}
		\hline {} &\multicolumn{4}{|c|}{\bf Explicit Solution} \\
		\hline {} &\multicolumn{2}{|c|}{\bf Trapezoidal} & \bf Simpson's ${1}/{3}$ & \bf Simpson's ${3}/{8}$\\
		\hline \bf M &$1$& $6$ & $6$ & $6$ \\
		\hline \bf RMS &$3.62901$& $2.89821$ & $2.91712$ & $3.08562$ \\
		\hline \bf Time &$0.0054$& $0.012$ & $0.01$ & $0.014$ \\
		\hline
	\end{tabular} 
\caption{RMS error from ground truth and execution time for Explicit Solution simulation. The RMS error is also for all 10 seconds.  M indicates the number of subintervals used per second in the numeric integration.  The numbers are also averaged over 20,000 seeds.}
	\label{table:Explicit solution}
\end{table}
It is clear that our Explicit Heston method is more accurate and quicker than the
other methods.
However, to get a single measure of improvement, we combine performance and time factors
and define
\begin{equation}\label{ExplicitGain}
	\text{Explicit Gain}=\frac{\tau_{\text{Other}}}{\tau_{\text{Explcit}}},
\end{equation}
where $\tau_{\text{Explcit}}$ and $\tau_{\text{Other}}$ are the execution times for
our Explicit Heston algorithm and some other method for a \emph{fixed} performance.
However, it is very hard to get the Milstein method, let alone the Euler one,
to perform as well as the worst we can 
do with the explicit weak solution method so we plot existing Milstein points and extend a smooth curve
to get some estimates.
(Part of the difficulty of collecting Milstein data with more steps here is that we would 
have to re-run the ground truth
with a much higher number of steps, which would exceed our computational limits.)
In this way, we estimate it would take Milstein at least $5.9$ s with a very high number of
steps to match the Explicit's $3.62901$ RMS
so the explicit gain in execution time would be $1093$.
We follow a similar procedure for Euler and tabulate the gains in Table \ref{table:Explicit gain}.
\begin{table}[H]
	\centering
	\begin{tabular}{|c|c|c|}
		\hline \bf Method &Euler & Milstein  \\
		\hline \bf Explicit Gain over &$2630$& $1093$ \\

		\hline
	\end{tabular} 
	\caption{Explicit Gain over Euler and Milstein.  These numbers
	indicate the number of times faster the explicit simulation is
	over Euler and Milstein for the same accuracy over the 10 s simulation.}
	\label{table:Explicit gain}
\end{table}
Clearly there is significant gain in using our Explicit simulations.
There are similar gains (exceeding $1000$) at other error levels and durations $T$.


\section{Performance of SA and Heston Algorithms} \label{Comparison}

Now, we turn our attention to option pricing.
For simplicity, we will use the same bases functions for volatility, price
and, in the case of Asian options, average price.
This means we will use $J=j^2$ (or $J=j^3$ in the case of Asian options)
functions of the form $e(s,v)=e_{k_1}(s)e_{k_2}(v)$
for $k_1,k_2\in\{1,...,j\}$.
Moreover, since there was little difference between Trapezoidal,
Simpson's $1/3$ and Simpson's $3/8$ in the simulation experiment
above, we will only consider the Trapezoidal method within our Heston
algorithms to follow.

\subsection{Weighted Heston on American Puts with LSM Algorithm}
First, we compare our Weighted Heston algorithm with the traditional Euler and Milstein
methods in pricing an American put option.
It was shown in the previous section that Explicit Heston simulation
is three orders of magnitude faster (for the same accuracy) as
Euler and Milstein simulation.
Now, we consider the real problem of option pricing and answer
the question: ``Does much does faster simulation translate into significantly
faster option pricing where, in addition to simulation, one has to
do dynamic programming to price?"
In addition, we do not assume the explicit case where Condition (C)
holds, which means the likelihoods must be computed.
For clarity, we do not use our SA algorithm yet, but rather stick to the LSM
algorithm.
We simply substitute our Weighted Heston as well as the other methods into
the simulation portion of this popular algorithm.

We use Heston and American put option parameters:
$\nu={8.1\kappa^2}/{4},\mu=0.0319,\rho=-0.7,\varrho=6.21,\kappa=0.2, S_0=100,V_0=0.501, T=50$ and the strike price $K=100$.
Here $n=8.1\notin\mathbb N$ and Condition (C) does not hold.
Hence, we use the full Weighted Heston algorithm with $\nu_\kappa=2\kappa^2$ in the closest
explicit Heston model. 
Finally, we use the weighted Laguerre polynomials  
\begin{eqnarray}
	e_1(x)&=&L_0(x)=exp(-x/2)\\
	e_2(x)&=&L_1(x)=exp(-x/2)(1-x)\\
	e_3(x)&=&L_2(x)=exp(-x/2)(1-2x+\frac{x^2}{2})\\
	e_{j}(x)&=&L_{j-1}(x)=exp(-x/2)\frac{e^x}{(j-1)n!} \frac{d^j}{dx^j}(x^je^{-x})
\end{eqnarray} 
with $j=3,\ J=3^2$ for the LSM pricing process.

Pre-experiments show that all these methods work and converge to the same \emph{nearly} correct
answer as the number of particles increases and the step size decreases.
The fact that they do not converge to the correct answer is due to the finiteness
of the collection of functions $\{e_k\}_{k=1}^j$ used.
Hence for a ground truth, we run the LSM algorithm with Milstein simulation with
extraordinarily fine time step and an enormous number of particles but still
for small $j=3$ (so the LSM algorithm can even work). 
(We will get around this small $j$ issue later when using SA instead of LSM.)
Table \ref{table:Pricing_Ground Truth for Weighted AP} gives the ground truth using a million particles with  $\Delta t=1/M=1/1,000$.

	\begin{table}[H]
		\centering		
		\begin{tabular}{|c|c|}
			\hline \multicolumn{2}{|c|}{ \bf Ground Truth}\\
			\hline \bf N & $1,000,000$\\
			\hline \bf M & $1,000$\\
			\hline \bf Option Price& $12.269$\\
			\hline
		\end{tabular} 
		\caption{Best estimate of American Put fair price.  This
		price no longer changes by increasing N or M.}
		\label{table:Pricing_Ground Truth for Weighted AP}
	\end{table}	

To compare performance, we will fix the error for the three methods and compare their execution time. 
The error is defined as:
\begin{equation}
	\emph{error}=\frac1{\text{\# Seeds}}\sum_{\text{Seeds}}\mid P^E-P\mid
\end{equation}
with $P^E$ being the option price obtained by running $N$ particles with Euler scheme and $P$ being the ground truth option price (except $J=3^2$ still). 
The other error are defined similarly. 
The results are provided in Tables \ref{table:WeightedsolutionAP} and \ref{table:Weighted solutionAP2}
for the cases where we can tolerate a pricing error of $4$ and $3$ cents respectively.
\begin{table}[H]
	\centering
	\begin{tabular}{|c|c|c|c|}
		\hline {} &\bf Euler &\bf Milstein& \bf Weighted Heston\\
		\hline \bf N & $10,000$ & $7,225$ & $2,500$ \\
		\hline \bf M & $100$ & $85$ & $15$\\
		\hline \bf Price & $12.3116$ & $12.2254$ & $12.2258$ \\
		\hline \bf Error & $0.0426$ & $0.0436$ & $0.0432$\\
		\hline \bf Time & $17.4178$ & $13.156$ & $1.387$ \\
		\hline \bf Time Gain & $1$ & $1.324$ & $12.562$ \\
		\hline
	\end{tabular} 
	\caption{American Put Execution Time - Low Accuracy case.  
	This is an execution time comparison to come within $4.3$ cents of the
	true price.  In each case, the best $N,M$ pairing was chosen to
	minimize time for this accuracy.}
	\label{table:WeightedsolutionAP}
\end{table}
	
\begin{table}[H]
	\centering
	\begin{tabular}{|c|c|c|c|}
		\hline {} &\bf Euler &\bf Milstein& \bf Weighted Heston\\
		\hline \bf N & $40,000$ & $30,625$ & $3,500$  \\
		\hline \bf M & $200$ & $175$ & $17$ \\
		\hline \bf Price & $12.3013$  & $12.2367$ & $12.2366$ \\
		\hline \bf Error & $0.0323$ & $0.0323$ & $0.0324$ \\
		\hline \bf Time & $143.356$ & $84.6254$  & $2.20966$\\
		\hline \bf Time Gain& $1$ & $1.694$  & $64.877$ \\
		\hline
	\end{tabular} 
	\caption{American Put Execution Time - High Accuracy case.
	This is an execution time comparison to come within $3.2$ cents of the
	true price.}
	\label{table:Weighted solutionAP2}
\end{table}
In Tables \ref{table:WeightedsolutionAP} and \ref{table:Weighted solutionAP2}, 
we defined a 
\begin{equation}\label{TimeGain}
\text{Time Gain}=\frac{\tau_{\text{LSM-Euler}}}{\tau_{\text{Other}}},
\end{equation}
where $\tau_{\text{LSM-Euler}}$ is the time required to achieve a specified
accuracy using the LSM algorithm with Euler simulation and $\tau_{\text{Other}}$
is the time required to obtain the same level of accuracy with some other method.
This resembles the Explicit Gain in (\ref{ExplicitGain}). 
Since in this experiment only the LSM is used,
Time Gain here describes how many times faster option pricing with the Milstein and Weighted Heston algorithms 
are than the basic Euler Scheme with the same error. 
As presented above, the weighted Heston algorithm shows a remarkable 
improvement over the traditional discretization method. 
The speed advantage is more significant when we require a higher accuracy.
Later, we will replace the LSM with the SA algorithm to increase speed further
and to enjoy the higher accuracy afforded by larger $J$.
	
\subsection{Weighted Heston on Asian Straddles with LSM Algorithm}
	
We compare Euler, Milstein and our weighted Heston by pricing Asian Straddles via the LSM algorithm. 
The discounted payoff process for an Asian straddle is $Z_t=e^{-\mu t}|R_t-K|$, where $R$ is the running
average of the Heston price, calculated as
\begin{equation}
	R_t=\frac{t-1}{t}R_{t-1}+\frac1tS_t,
\end{equation}
and $K$ is the strike price. 
As the Asian Straddles option pricing model is a three factor model
(spot price, average price and volatility), we will only use $j=2$ for each 
factor for computational reasons. 
The other parameters remain the same as the American put option: $\nu={8.1\kappa^2}/{4},\mu=0.0319,\rho=-0.7,\varrho=6.21,\kappa=0.2, S_0=100,V_0=0.501, T=50$ and the strike price $K=100$. 
The groudtruth of the Asian Straddles price, computed by Milstein's method with a million particles and a very fine 
time step, is used for measuring the error and is given in Table \ref{table:Pricing_Ground Truth for Weighted AS}. 
\begin{table}[H]
	\centering		
	\begin{tabular}{|c|c|}
		\hline \multicolumn{2}{|c|}{ \bf Ground Truth}\\
		\hline \bf N & $1,000,000$\\
		\hline \bf M & $1,000$\\
		\hline \bf Option Price& $136.174$\\
		\hline
	\end{tabular} 
	\caption{Best estimate of Asian Straddle fair price.  This
		price no longer changes by increasing N or M. }
	\label{table:Pricing_Ground Truth for Weighted AS}
\end{table}

The Asian straddle time gains, given in Tables \ref{table:Weighted solutionAS} and \ref{table:Weighted solutionAS2} (to follow),
also indicate the efficiency of the weighted Heston as it did for the American put. 
\begin{table}[H]
	\centering
	\begin{tabular}{|c|c|c|c|}
		\hline {} &\bf Euler &\bf Milstein &\bf Weighted Heston\\
		\hline \bf N & $10,000$ & $4,900$ & $3,510$  \\
		\hline \bf M & $100$ & $70$ & $12$  \\
		\hline \bf Price & $135.956$ & $135.952$ & $136.019$ \\
		\hline \bf Error & $0.218$ & $0.214$ & $0.222$ \\
		\hline \bf Time & $18.8237$ & $11.2313$ & $1.8943$ \\
		\hline \bf Time Gain & $1$ & $1.676$ & $9.937$ \\
		\hline
	\end{tabular} 
	\caption{Asian Straddle Execution Time - Low Accuracy case.
	This is an execution time comparison to come within $22$ cents of the
	true price.}
	\label{table:Weighted solutionAS}
\end{table}
For lower accuracy, the weighted Heston performs about ten times as fast as the traditional method with the fixed error. 
As with the American put, this outperformance improves as one desires higher accuracy.
\begin{table}[H]
	\centering
	\begin{tabular}{|c|c|c|c|}
		\hline {} &\bf Euler &\bf Milstein &\bf Weighted Heston\\
		\hline \bf N & $40,000$ & $25,600$ & $4,800$  \\
		\hline \bf M & $200$ & $160$ & $13$ \\
		\hline \bf Price & $136.043$ & $136.046$ & $136.303$ \\
		\hline \bf Error & $0.131$ & $0.128$ & $0.124$ \\
		\hline \bf Time & $145.864$ & $73.958$ & $2.861$ \\
		\hline \bf Time Gain & $1$ & $1.972$ & $50.984$ \\
		\hline
	\end{tabular} 
	\caption{Asian Straddle Execution Time - High Accuracy case.
	This is an execution time comparison to come within $13$ cents of the
	true price.}
	\label{table:Weighted solutionAS2}
\end{table}
Our weighted Heston method shows a rather strong performance in the high accuracy case
since the Time Gain increases to around $51$, which means we can get the same accuracy with ${1}/{51}$ the execution time.
Indeed, these results show that the simulation component of the LSM algorithm
is very important and that our Weighted Heston method is the best method.
	
We can speculate on the reason the outperformance is less for the Asian straddle than the American put:
The method and time in going from spot price to running average price is the same, whether we use
Euler, Milstein or Weighted Heston.
Moreover, adding a constant (running average price time) to the numerator and denominator of
(\ref{TimeGain}) will drag the Time Gain ratio towards $1$.
	
\subsection{Comparison of SA and LSM on American Puts}

Having shown that our Explicit and Weighted
Heston simulation methods can be superior to the Euler and Milstein methods in option pricing, 
we turn our attention to comparing 
the SA and LSM algorithms with different numbers and types of functions $\{e_k\}_{k=1}^J$ used.
In this subsection, we will use model parameters: 
\(
\mu=0.0319,\rho=-0.7,\varrho=6.21,\kappa=0.61, K=100, S_{0}=100, V_{0}=0.0102, T=50
\)
and $\nu=\kappa^2/2$ so the Explicit algorithm applies. 
We use $\gamma=2.115,0.195,0.0095$ for $J=2^2,3^2,4^2$ respectively in the case $N=10,000$ and
$\gamma=1.068,0.762,0.0082$ for $J=2^2,3^2,4^2$ respectively in the case $N=100,000$ below
as these were determined numerically to be reasonable choices.
All the prices are calculated by taking the average of $100$ independent experiment.

First, we show that the LSM algorithm can fail numerically when adding more weighted 
Laguerre functions in an attempt to achieve higher price accuracy.
Tables \ref{table:Pricing1}, \ref{table:Pricing2} show this along with performance.
\begin{table}[H]
\centering
\begin{tabular}{|c|c|c|c|c|}
	\hline {} & \bf SA Price &\bf SA Time & \bf LSM Price& \bf LSM Time\\
	\hline \bf J=$2^2$& $8.44858$ &$0.11298$& $8.40775$& $0.124679$\\
	\hline \bf J=$4^2$ & $8.49936$ &$0.14411$ & $8.38028$ & $0.258755$\\
	\hline \bf J=$8^2$& $8.41892$ & $0.2566856$ &$5.58625$ & $2.13897$\\
	\hline
\end{tabular}
\caption{American Put price obtained using SA and LSM with $N=10,000$ particles.  Increasing the number of bases functions $J$ should give better estimates of the $\$8.59$ fair price but there is numerical instability for LSM.}
\label{table:Pricing1}
\end{table}

\begin{table}[H]
	\centering		
	\begin{tabular}{|c|c|c|c|c|}
		\hline {} & \bf SA Price &\bf SA Time & \bf LSM Price& \bf LSM Time\\
		\hline \bf J=$2^2$& $8.4213$ &$1.24712$& $8.39404$& $1.51143$\\
		\hline \bf J=$4^2$ & $8.50788$ &$1.79924$ & $8.51376$ & $2.7524$\\
		\hline \bf J=$8^2$& $8.51644$ & $2.64996$ &$7.18587$ & $20.1488$\\
		\hline
	\end{tabular} 
	\caption{American Put price obtained using SA and LSM with $N=100,000$ particles.  Increasing the number of bases functions $J$ should give better estimates of the $\$8.59$ fair price but there is numerical instability for LSM.}
	\label{table:Pricing2}
\end{table}
We can draw several conclusions from Tables \ref{table:Pricing1} and \ref{table:Pricing2}. 
First, there is a large execution time advantage for our SA algorithm
over the popular LSM algorithm, especially as $J$ increases and matrix
inversion becomes difficult.
For small numbers of the basis functions, SA is about  $10\%$ faster than LSM. 
However, when the number of basis functions increases, the SA time performance 
becomes even more superior. 
For example, when $J=8^2$, the SA algorithm is nearly ten times faster,
yet much more accurate.
Next, given \emph{enough} particles (eg. $N=100,000$ here), prices and pricing accuracy should 
both increase as we add more basis functions because we will obtain a better estimate
of the optimal stopping time. 
Table \ref{table:Pricing2} does demonstrate that as $J$ increases from $2^2$ to $8^2$ 
the SA option prices increase and the SA algorithm does not break.
Indeed, it should never break as it avoids the numeric issues of matrix
inversion.
The LSM algorithm does break as prices dive and time spikes for large $J$
in both Table \ref{table:Pricing1} and Table \ref{table:Pricing2} due
to ill-conditioned matrix inversion in the least squares estimate.
Prices fall in Table \ref{table:Pricing1} for the SA algorithm for a
different reason:
When $N$ is small the projection parameter estimates are often bad,
especially when there are a lot of parameters to estimate, and
optimal stopping is easily missed, even when $J$ is large.
More bad (low $N$) parameter estimates with larger $J$ is not necessarily an advantage
and prices can vary in either direction as you increase $J$ with
small $N$ fixed.
To provide further evidence of this expected price improvement in $J$ 
given large enough $N$ and to find the ground truth for pricing, we also run the 
Stochastic Approximation method with $N=1,000,000$ and $J=12^2$. 
As shown in Table \ref{table:Pricing_Ground Truth}, 
the American put option price rises to 8.58712.
\begin{table}[H]
	\centering		
	\begin{tabular}{|c|c|}
		\hline \multicolumn{2}{|c|}{ \bf Ground Truth}\\
		\hline \bf N & $1,000,000$\\
		\hline \bf J & $12^2$\\
		\hline \bf $\gamma$ & $0.99294$\\
		\hline \bf SA Option Price& $8.58712$\\
		\hline
	\end{tabular} 
	\caption{Best estimate of American Put fair price.  This
		price is obtained by the SA method and no longer changes by increasing N or J.}
	\label{table:Pricing_Ground Truth}
\end{table}
The SA prices in Tables \ref{table:Pricing1} and \ref{table:Pricing2}
were heading in the right direction.
The SA algorithm behaves better than the LSM, especially as the desired accuracy increases.

\subsection{Comparison of SA and LSM on Asian Calls}

We continue our comparison of SA and LSM algorithms but now on an Asian Call
option and in a situation where the Weighted Heston has to be used.
First an observation: Since we are pricing options on average spot price in 
Asian options, which varies less and less as time goes on, the pricing
problem should be easy.
Suppose we are slightly off on our optimal stopping time and the optimal
stopping time is not near the beginning of the period.
Then, the average price and the payoff will not differ much between the
optimal stopping time and our estimate (due to the averaging) and hence
our price estimate and the optimal option price will not either.

In this section, we will use model parameters: $\nu={8.1\kappa^2}/{4},  \mu=0.0319,\rho=-0.7,\varrho=6.21,\kappa=0.2 $ and $T=50$ so $n=8.1$ and $\nu_\kappa=2\kappa^2$ is used in the Closest Explicit Heston. 
The ground truth for this experiment is:
\begin{table}[H]
	\centering		
	\begin{tabular}{|c|c|}
		\hline \multicolumn{2}{|c|}{ \bf Ground Truth}\\
		\hline \bf N & $1,000,000$\\
		\hline \bf J & $12^3$\\
		\hline \bf $\gamma$ & $0.962$\\
		\hline \bf SA Option Price& $31.3455$\\
		\hline
	\end{tabular} 
	\caption{Best estimate of  Asian Call fair price.  This
		price is obtained by the SA method and no longer changes by increasing N or J.}
	\label{table:Pricing_Ground Truth}
\end{table}
Again, it is impossible to get that accurate on a standard
contemporary computer with the LSM method due
to matrix inversion issues for large $J$.
Also, Euler and Milstein would not finish within a two week time frame
for this value of $N$ and a high enough number of steps $M$.
All the prices are calculated by taking the average of 100 independent experiments.

Following the same procedure as pricing the American Put option, 
we first consider performance with different numbers of basis functions and
show this in Table \ref{table:Asian Pricing}:

\begin{table}[H]
	\centering		
	\begin{tabular}{|c|c|c|c|c|}
		\hline {} & \bf SA Price &\bf SA Time & \bf LSM Price& \bf LSM Time\\
		\hline \bf N & \multicolumn{2}{|c|}{$100,000$}&\multicolumn{2}{|c|}{$100,000$} \\
		\hline \bf J=$2^3$& $31.3411$ &$11.2404$& $25.2365$& $12.511$\\
		\hline \bf J=$4^3$& $31.3411$ & $36.2066$ &$20.3398$ & $92.432$\\
		\hline
	\end{tabular} 
	\caption{Asian Call price obtained using SA and LSM with $N=100,000$ particles.  Increasing the number of bases functions $J$ should give better estimates of the $\$31.345$ fair price but there is numerical instability for LSM.}
	\label{table:Asian Pricing}
\end{table}
For completeness, we used $\gamma=1,\,0.824$ for $J=2^3,4^3$ respectively.

We can clearly see that the LSM fails already when $J=2^3$. 
The main reason still lies in the matrix inversion part: 
Since the Asian Calls is a three factor model, we have to invert a
$8\times 8$ matrix.
Indeed, when you have both price and average price there is a greater
chance of this matrix having nearly linearly dependent rows and
hence being highly ill-conditioned to inversion.

The SA algorithm does not fail even for large numbers of basis functions. 
The price remains the same for $J=2^3$ and $4^3$ due to the averaging mentioned
in the first paragraph above.
Indeed, a comparison between Tables \ref{table:Pricing_Ground Truth} and \ref{table:Asian Pricing} shows that the SA algorithm with $J=2^3,4^3$ and $N=100,000$ already gives a rather close result to the ground truth.
	
\subsection{Comparison of Weighted-SA and Euler-LSM on American Puts}
Our final results are comprehensive, showing the overall gain of the methods suggested
herein over the traditional Euler-LSM method.
The model parameters used in this section are: $\nu={8.1\kappa^2}/{4},\mu=0.0319,\rho=-0.7,\varrho=6.21,\kappa=0.2 $ and $T=50$ 
so $n=8.1\notin\mathbb N$ and Condition (C) does not hold.
Hence, we will use the full Weighted Heston algorithm with $\nu_\kappa=2\kappa^2$ in the closest
explicit Heston model. The initial state $S_0=100, V_0=0.102$, and the strike price $K=100$.
	
The ground truth price is found using the weighted Heston in SA algorithm with fine meshing.
The result is given in Table \ref{table:Pricing_Ground Truth WSA}.
\begin{table}[H]
	\centering		
	\begin{tabular}{|c|c|}
		\hline \multicolumn{2}{|c|}{ \bf Ground Truth}\\
		\hline \bf M & $5$\\
		\hline \bf N & $1,000,000$\\
		\hline \bf J & $12^2$\\
		\hline \bf $\gamma$ & $0.00628$\\
		\hline \bf SA option price& $7.9426$\\
		\hline
	\end{tabular} 
	\caption{Best estimate of American Put fair price.  This
		price is obtained by the SA method and no longer changes by increasing M, N or J.}
	\label{table:Pricing_Ground Truth WSA}
\end{table}
	
We run the actual experiment by varying $M,N,J$ to obtain the option price for fixed execution times. 
\begin{table}[H]
	\centering
	\begin{tabular}{|c|c|c|c|c|}
		\hline {} & \bf E-LSM& \bf  W-SA& \bf E-LSM& \bf  W-SA\\
		\hline \bf M & $100$& $5$& $100$& $5$\\
		\hline \bf N & $10,000$ & $65,000$& $10,000$ & $90,000$\\
		\hline \bf J &$4^2$ & $8^2$ & $5^2$ & $6^2$\\
		\hline \bf price &$7.371$ & $7.932$ & $6.944$& $7.9347$\\
		\hline \bf error & $0.572$ & $0.0103$&$0.9986$ & $0.00788$\\
		\hline \bf time &$19.662$ & $19.433$& $22.702$ & $22.528$\\
		\hline \bf performance gain& $1$& $55.534$& $1$& $126.726$\\
		\hline
	\end{tabular}
	\caption{Combined RMS Pricing comparison on American Puts for fixed execution time.  Due to numeric instability of the Euler-LSM method, the performance gain can becomes arbitrarily large as the need for accuracy increases.  In all cases, near optimal $M,N,J$ were used.}
	\label{table:Final Pricing Comparison}
\end{table}
(For clarity, $\gamma$ was taken as $0.00096$ and $0.013$ in the $N=65,000$ and $90,000$
cases respectively.)

The Performance Gain is defined (in a similar way as the time factor 
in the previous section) to represent the relative accuracy of each method given a fixed computation time. 
The traditional Euler-LSM method does not fail in $J=4^2$ case as is shown
in the first column.
In this situation, the accuracy will be increased by $55$ times by switching to the Weighted-SA method. 
The last two columns present the case that Euler-LSM starts to fail. 
As we will not know the ground truth, hence if the LSM is failing in practice, it is still resonable to conduct the comparison in this case. 
We found that the relative accuracy has risen to more than $126$ times using the new algorithms, which is an impressive 
\emph{two-orders of magnitude} improvement for pricing options in the real market.
We mention in future work below ways to increase this even more.


\section{Conclusion} \label{Conclusions}

We can make the following conclusions:
(1) The Heston model has explicit weak stochastic differential 
	equation solutions.  These solutions can be easily constructed
	when Condition (C) holds. Otherwise, they have an explicit likelihood
	that can be used either as a weight or to change probabilities so
	the desired model holds.
(2) The Explicit Heston algorithm should be considered for simulation when
	it applies.  In particular, it does not produce negative volatility
	values and it compares favourably in terms of both performance and
	execution time to the Euler and Milstein methods.
	Indeed, we showed a three order of magnitude overall advantage.
(3) The Weighted (or Explicit when it applies) Heston algorithm should be
	considered for Monte Carlo option prices.
	It compares favorably to the Euler and Milstein methods
	on the American and Asian option pricing examples considered herein.
	(It is also much easier to implement than the Broadie-Kaya method
	on path-dependent options.)
(4) Stochastic Approximation (SA) should be considered as a favorable alternative
	to Least-Squares regression in the LSM algorithm.
	It avoids numerically nasty matrix inversion and thereby allows a larger number
	$J$ of functions in the projection and closer approximation
	of the future payoff conditional expectations.

Potential future work includes:
(1) The SA pricing algorithm should be explored more.
Are the situations where the LSM algorithm should still be used?
Will other stochastic approximation schemes yield better performance?
Are there any guidelines for selecting the functions $(e_k)$?	
(2) The Explicit and Weighted Heston algorithms need to be explored more.
What type of numeric integration is best?	
Are there variations of the algorithm that perform better?
(3) Resampling could be employed to improve the performance of the Weighted Heston algorithm.
Currently, we keep all paths, including those that have very low weight.
It may be a better strategy to split the higher weight ones and remove the lower weight
ones in an unbiased way.
However, this must be done in the correct way since American and Asian option pricing are
path dependent problems.
It will not be enough to just worry about the current particle states.
We will have to consider the whole particle paths.
(4) Precise conditions for rate of convergence results and the optimal
rates should be found for the combined Weighted Heston SA algorithm.
This is not necessarily simple because of the weak interaction and
the path-dependence.
(5) New explicit weak solutions to other financial models should be investigated.
The author is very optimistic that there are
explicit three-factor stochastic-mean, stochastic-volatility models for the finding.
This would be done along the lines laid out in the appendix.


\section{Appendix: Solving the SDEs} \label{SolveSDEs}

\subsection{Background}
Generally, a weak solution (on a subdomain of $\mathbb R^p$) to 
\begin{equation}
dX_{t}=b(X_{t})dt+\sigma (X_{t})dW_{t}  \label{sdeX}
\end{equation}
is the triplet of a filtered probability space $(\Omega ,{{%
\mathcal{F}}},\{{\mathcal{F}}\}_{t\geq 0},P)$, a $\mathbb{R}^{d}$-valued
Brownian motion $\{W_{t},\ t\geq 0\}$ with respect to $\{{\mathcal{F}}%
_{t}\}_{t\geq 0}$, and an $\{{\mathcal{F}}_{t}\}_{t\geq 0}$-adapted
continuous process $\{X_{t},\ t\geq 0\}$ such that $(W,X)$ satisfy Equation (%
\ref{sdeX}). 
More restrictively, a strong solution to (\ref{sdeX}) is
an \{$\mathcal{F}_{t}^{W}\}_{t\geq 0}$-adapted process $X$
on a probability space $(\Omega ,{{\mathcal{F}}},P)$ supporting
the Brownian motion $W$, where $\mathcal{F}_{t}^{W}\circeq \sigma
\{W_{u},\,u\leq t\}$.

Weak solutions are often handled via martingale problems: Suppose $D\subset \mathbb R^p$
is a domain, $C_D[0,\infty)$ denotes the continuous $ D$-valued functions
on $[0,\infty)$ with the topology of uniform convergence on compacts, 
$(L,\mathcal{D}(L))$ is a linear operator on $C(D)$,
the continuous $\mathbb R$-valued functions on $D$,
and $\mu$ is a probability measure on $D$. 
Then, a solution to the $C_D[0,\infty)$-martingale problem for $(L,\mu)$ is any
probability measure $P_{\mu}$ on $\Omega=C_D[0,\infty)$ such that the canonical
process $\{\omega_t,\ t\ge 0\}$ satisfies: 
$P_{\mu}\omega_{0}^{-1}=\mu$, and for each $f\in \mathcal{D}(L)$ one has
that 
\begin{equation}
M_{t}(f)(\omega)=f(\omega _{t})-\int_{0}^{t}Lf(\omega _{u})du,\quad t\geq 0, 
\end{equation}
is a $P_{\mu}$-martingale. 
The martingale problem is well-posed
if there is exactly one such probability measure $P_{\mu}$ on $C_D[0,\infty)$.

A weak solution $((\Omega ,{{\mathcal{F}}},\{{\mathcal{F}}_{t}\}_{t\geq
0},P),\{W_{t},\ t\geq 0\},\{X_{t},\ t\geq 0\})$ to (\ref{sdeX}) then (see
Karatzas \& Shreve 1987 p. 317) corresponds to each martingale problem
solution $P_{\mu}$ for $(L,\mu)$, with $L$ defined by 
\begin{equation}
Lf(x) =\sum_{i=1}^{p}b_{i}(x)\partial
_{x_{i}}f(x)  \label{Ldef} 
+\frac{1}{2}\sum_{i=1}^{p}\sum_{j=1}^{p}a_{ij}(x)\partial
_{x_{i}}\partial _{x_{j}}f(x),  
\end{equation}
through the relation $(\Omega ,{{\mathcal{F}}})=(C_D[0,\infty ),\mathcal{B}%
(C_D[0,\infty )))$, $X_{t}=\omega _{t}$ for $t\geq 0$, $P_{\mu}=PX^{-1}$ ,
where $\omega _{t}$ denotes the projection function on $C_D[0,\infty )$. 
$(W_{t},{%
\mathcal{F}}_{t})_{t\geq 0}$ are defined through a martingale representation
theorem and $a=\sigma \sigma ^{T}$, where $\sigma \in \mathbb{R}^{p\times
d} $.
Well-posedness of a martingale problem is with respect to the given operator $L$
(and initial distribution $\mu$).
It opens the possibilities of having different SDEs with the same operator and hence
(under well-posedness) the same law.
We will take advantage of this fact in (\ref{ExtendHesa},\ref{ExtendHes}) below.

The Heston model (\ref{Heston}) corresponds to the martingale problem for operator 
\begin{eqnarray}
Lf(s,v) &\!\!=&\!\!\mu s\,\partial_{s}f(s,v)+(\nu-\varrho v) \partial_{v}f(s,v)  \label{LHestdef} 
+\frac{1}{2}s^2v\,\partial^2_sf(s,v)\\\nonumber
&\!\!+&\!\!\rho\kappa sv\,\partial_{s}\partial _{v}f(s,v)
+\frac{1}{2}\kappa^2v\,\partial^2_vf(s,v).  
\end{eqnarray}
However, $b$ and $\sigma$ are not bounded
nor is $a=\sigma\sigma'$ is strictly positive definite everywhere.
Hence, well-posedness of this martingale problem is not immediate.
However, it follows from the proofs in \citet{StroockVaradhan:1969}, \citet{StroockVaradhan:1979}
that there is uniqueness up to the first time the volatility hits zero.
This means that there is well-posedness in the case $\nu\ge {\kappa^2}/2$ since
it is well known that the (CIR) volatility will not hit zero in this case and we have
already discussed existence.
As for the remaining case, we mention that 
\citet{DaskalopoulosFeehan} and others have recognized the degenerate nature of the
Heston model and considered a different type of existence and uniqueness.

Our work gives explicit construction of the weak solutions that are known to be
distributionally unique in the case $\nu\ge {\kappa^2}/2$. 
Its importance is in the ability to simulate these explicit constructions.
Moreover, our methods may well yield explicit solutions for other financial models.

\subsection{Proof of Theorem \ref{Theorem1}}\label{Theorem1Proof}

Stochastic differential equations can be interpreted and solved explicitly
either in the strong or weak sense. 
Weak interpretations are often
sufficient in applications like mathematical finance and filtering and allow
solutions to a greater number of equations than strong solutions.
However, there is also the possibility of finding new explicit strong
solutions through the guise of weak solutions, which should not
be surprising given the result of \citet{Heunis}.
Moreover, weak solutions can often be converted to (marginals of) strong
solutions of a higher dimension SDE, which is the first way that we will use
weak interpretations.
Our approach will be to show everything explicitly in the case $n=2$ and then
explain the necessary changes for $n\in\{1,3,4,...\}$.
However, we first simplify the task by observing the ``independently driven''
part of the price can be split off.
\subsubsection{Price Splitting}
Suppose that 
\begin{eqnarray}
d\left(\begin{array}{c}S^c_{t}\\V_{t}\end{array}\right)\label{CorrHeston}
&=&\left(\begin{array}{c}\mu  S^c_{t}\\\nu -\varrho  V_{t}\end{array}\right)dt+\left(\begin{array}{c}\rho S^c_{t} V_{t}^{\frac{1}{2}}\\\kappa  V_{t}^{\frac{1}{2}}\end{array}\right)d\widehat \beta_{t},\\\label{IndHeston}
S^i_t&=&\exp\left(\sqrt{1-\rho^2}\int_0^t  V_s^\frac12 dB_s-\frac{1-\rho^2}2\int_0^t  V_sds\right)
\end{eqnarray}
with respect to independent Brownian motions $\widehat \beta,B$.
Then, it follows by It\^{o}'s formula and the independence of $\widehat \beta,B$
that $ S_t=S^c_t S^i_t$ and $ V_t$ satisfy (\ref{Heston}) with $\beta=\hat\beta$.
Moreover, $S^i$ is conditionally (given $ V$) log-normal and hence trivial to simulate.
Hence, we only have to solve (\ref{CorrHeston}), which we do using weak interpretations
to create a higher dimension SDE
that does satisfy (\ref{Bracket1}) and hence has an \emph{explicit} strong solution.

\subsubsection{Volatility in Case $n=2$}\label{Volatility2}

To ease the notation, we will use $Y$ and $Z$ in place of $Y^1,Y^2$ in Theorem \ref{Theorem1}.
We consider solutions to a Cox-Ingersoll-Ross (CIR) type Ito equation 
\begin{equation}\label{CIRModel}
d V_{t}
=\left(\nu -\varrho  V_{t}\right) dt+\kappa \sqrt{ V_{t}}\,d\widehat \beta_{t}, 
\end{equation}
for some Brownian motion $\widehat \beta$. 
Let $W^1,W^2$ be
independent Brownian motions so 
\begin{equation}\label{BasicYZ}
Y_t=\frac{\kappa}{2}\int_{0}^{t}e^{-\frac{\varrho}2 (t-u)}dW^1_u+e^{-\frac{\varrho}2t}Y_0,\
Z_t=\frac{\kappa}{2}\int_{0}^{t}e^{-\frac{\varrho}2 (t-u)}dW^2_u+e^{-\frac{\varrho}2t}Z_0
\end{equation}
are independent Ornstein-Uhlenbech processes. 
It follows by It\^{o}'s formula
that, \emph{if} Condition (C) is true (with $n=2$),
then $ V=Y^{2}+Z^2$ satisfies 
(\ref{CIRModel}) with 
\begin{equation}\label{Bdefine}\widehat 
\beta_{t}=\int_{0}^{t}\frac{Y_u}{\sqrt{Y_{u}^2+Z^2_u}}dW^{1}_u 
+\int_{0}^{t}\frac{Z_u}{\sqrt{Y_{u}^2+Z^2_u}}dW^{2}_u. 
\end{equation}
(Note that $(\widehat \beta,W)$ is a standard two dimensional Brownian motion, where 
\begin{equation}
W_t=
\int_{0}^{t}\frac{Z_u}{\sqrt{Y_u^2+Z^2_u}}dW^1_u
-
\int_{0}^{t}\frac{Y_u}{\sqrt{Y_{u}^2+Z^2_u}}dW^2_u,
\end{equation}
by Levy's characterization.)
We call $( V,\widehat \beta)$ a weak solution since the
definition of $\widehat \beta$ was part of the solution. 
$ V$ will also be a strong
solution if $ V_{t}$ is measurable with respect to $\mathcal{F}%
_{t}^{\widehat \beta}\circeq \sigma \{\widehat \beta_{u},\,u\leq t\}$. 
A strong
solution does not immediately follow from the Yamada-Watanabe theorem since
the conditions for pathwise uniqueness in e.g. Theorem IX.3.5 of 
\citet{RevuzYor:1999} can not immediately be validated. 
Moreover, explicit form in terms of only $\widehat \beta$ is unknown. 
(Example 3.4 of Kouritzin 2000
shows that it unrepresentable
in terms of a single Ornstein-Uhlenbeck processs.)
Regardless, it is unimportant to us if $V$ is a strong solution or not. 
(There is a famous example of H. Tanaka of a simple SDE
with weak but not strong solutions.)

\subsubsection{Extended Price Formulation in Case $n=2$}

Recall $W^1,W^2$ are independent standard Brownian motions,
set
\begin{equation}\label{ExtendHesSigma}
	\sigma(y,z,s)=\left[\begin{array}{cc}\frac{\kappa}{2}&0\\0&\frac{\kappa}{2}\\
\rho\,sy& \rho\,s z
\end{array}\right]
\end{equation}
and define a new SDE of the form:
\begin{equation}\label{ExtendHesa}
d\!\left[\begin{array}{c}Y_t\\Z_t\\S^c_{t}\end{array}\right]
=\left[\begin{array}{c}-\frac{\varrho}2 Y_t\\-\frac{\varrho}2 Z_t\\\mu  S^c_t
\end{array}\right]dt+\!\sigma(Y_t,Z_t,S^c_t)\left[\begin{array}{c}dW^1_{t}\\dW^2_{t}\end{array}\right].
\end{equation}
This equation has a unique strong solution.
Indeed, the first two rows immediately give strong uniqueness for $Y,Z$ and then
$S^c$ is uniquely solved as a stochastic exponential (see e.g. Protter 2004).
This solution can be rewritten as:
\begin{equation}\label{ExtendHes}
d\!\left[\begin{array}{c}Y_t\\Z_t\\S^c_{t}\end{array}\right]
=\left[\begin{array}{c}-\frac{\varrho\,Y_t}2 \\-\frac{\varrho\,Z_t}2 \\\mu  S^c_t\end{array}\right]dt+\!\left[\begin{array}{cc}
\frac{\frac{\kappa}{2}\ Z_t}{\sqrt{Y_t^2+Z_t^2}}&\frac{\frac{\kappa}{2}\ Y_t}{\sqrt{Y_t^2+Z_t^2}}\\
\frac{-\frac{\kappa}{2}\ Y_t}{\sqrt{Y_t^2+Z_t^2}}&\frac{\frac{\kappa}{2}\ Z_t}{\sqrt{Y_t^2+Z_t^2}}\\
0 &\rho S^c_{t} V_{t}^{\frac{1}{2}}\end{array}\right]
\left[\begin{array}{c}dW_{t}\\d\widehat \beta_{t}\end{array}\right],
\end{equation}
where 
\begin{equation}\label{newBMs}
\left[\begin{array}{c}dW_{t}\\d\widehat \beta_{t}\end{array}\right]
=\left[\begin{array}{cc}\frac{Z_t}{\sqrt{Y_t^2+Z_t^2}}&\frac{-Y_t}{\sqrt{Y_t^2+Z_t^2}}\\\frac{Y_t}{\sqrt{Y_t^2+Z_t^2}}&\frac{Z_t}{\sqrt{Y_t^2+Z_t^2}}\end{array}\right]
\left[\begin{array}{c}dW^1_{t}\\dW^2_{t}\end{array}\right].
\end{equation}
Now, the last row of (\ref{ExtendHes}) together with
(\ref{CorrHeston},\ref{IndHeston},\ref{CIRModel},\ref{BasicYZ},\ref{Bdefine}) show that $( S=S^iS^c, V=Y^2+Z^2)$ is
the Heston model with $\nu ={\kappa^2}/2$.
Moreover, (\ref{ExtendHesSigma}) does satisfy (\ref{Bracket1}) since
\begin{equation}\label{ExtendHesBracket}
(\nabla \sigma_1)\sigma_2=\left(\!\begin{array}{c}0\\0\\\rho^2\,s\,y\,z\end{array}\!\right) =(\nabla \sigma_2)\sigma_1
\end{equation}
so we will be able to look for simple explicit solutions.
Our extended Heston system (\ref{ExtendHesa}) can also be written as a Stratonovich equation:
\begin{eqnarray}\label{ExtendHesStrat}
\!\!\!\!\!\!\!\!\!d\!\left[\!\begin{array}{c}Y_t\\Z_t\\S^c_{t}\end{array}\!\right]
&\!\!\!=&\!\!\!\!\left[\!\begin{array}{c}-\frac{\varrho}2 Y_t\\-\frac{\varrho}2 Z_t\\\mu  S^c_t-\frac{\kappa\rho S^c_t}2-S^c_t\rho^2\frac{Y_t^2+Z_t^2}2\end{array}\right]\!dt
+\!\left[\!\begin{array}{cc}\frac{\kappa}{2}&0\\0&\frac{\kappa}{2}\\
\rho\, S^c_tY_t&
\rho\, S^c_t Z_t\end{array}\!\right]\bullet\!\left[\!\begin{array}{c}dW^1_{t}\\dW^2_{t}\end{array}\!\right]\!,
\end{eqnarray}
where the stochastic integral implied by the $\bullet$ is now interpretted in the Fisk-Stratonovich sense.
We define the full Fisk-Stratonovich drift coefficient to be:
\begin{eqnarray}\label{ExtendStrDrift}
h(y,z,s,v)=\left[\begin{array}{c}-\frac{\varrho}2 y\\-\frac{\varrho}2 z\\\mu  s-\frac{\kappa\rho s}2-s\rho^2\frac{y^2+z^2}2\end{array}\right].
\end{eqnarray}
\begin{RM}
\emph{
Reformulating the Heston equations into a higher dimensional equation so
that commutator conditions like (\ref{ExtendHesBracket}) are true and
explicit solutions exist is one of our main contributions.
It is believed that similar techniques can be used on some other interesting
financial models.
}
\end{RM}

\subsubsection{Explicit Solutions for Extended Heston in case $n=2$} \label{ExplicitHest}

We can solve for the possible strong solutions to (\ref{ExtendHes}).
The first step is to transform the equation to a simpler one using Theorem 2 of
\citet{Kouritzin/Remillard:2015}, restated here in the case $p=3$ and $d=r=2$ for convenience:

\begin{thm}\label{KoRe}
Let $D\subset\mathbb R^3$ be a bounded convex domain, 
$X_0$ be a random variable living in $D$, $W$ be an $\mathbb R^2$-valued standard Brownian motion
and $h:D\rightarrow\mathbb R^3$, $\sigma:D\rightarrow\mathbb R^{3\times 2}$ 
be twice continuously differentiable functions with $\sigma(X_0)$ having full rank and satisfying (\ref{Bracket1}).
Then, the Stratonovich SDE
\(
dX_t =h(X_t)dt +\sigma(X_t)\bullet dW_{t}
\)
has a solution  
\begin{equation}
X_t=\Lambda^{-1}\left(\begin{array}{c}\overline X_t\\\widehat X_t\end{array}\right)
\end{equation}
on $[0,\tau]$ for some stopping time $\tau>0$,
in terms of a simpler SDE
\begin{equation}
\!\!\left[\begin{array}{c}\overline X_t\\\widehat X_t\end{array}\right]
=\int_0^t \widehat h\left(\begin{array}{c}\overline X_s\\\widehat X_s\end{array}\right)ds
+\left(\begin{array}{cc} W_t\\0\end{array}\right)+\Lambda(X_0),\,
\text{ with } \widehat h(x)=(\nabla \Lambda h)\circ\Lambda^{-1}(x),	
\end{equation}
and a local diffeomorphism $\Lambda$ 
if and only if the simpler SDE has a solution
up to a stopping time at least as large as $\tau$.
Without loss of generality, the local diffeomorphism can have the form $\Lambda=\Lambda_2\circ\Lambda_1$ for any local diffeomorphisms
$\Lambda_1:D\rightarrow \mathbb R^3$ satisfying $\nabla \Lambda_1\sigma_1\circ\Lambda_1^{-1}(x)=e_1$
and $\Lambda_2:\Lambda_1(D)\rightarrow \mathbb R^3$ satisfying 
$\{\nabla \Lambda_2\nabla \Lambda_1\sigma_2\}\circ(\Lambda_1^{-1}\circ\Lambda_2^{-1}(x))=e_2$,
where $(e_1\, e_2\,e_3)=I_3$ is the identity matrix. 
\end{thm}	
There are four things to note:
	(1) 
The diffusion coefficient is just $\left( I_2, 0\right)'$ for the simpler
SDE.
In this case, there is no difference between the It\^{o} and Stratonovich equations so
we have just stated the simpler SDE as the more common It\^{o} equation.
(2)
We can check this local solution to see if it is actually a global
solution.
We will do this below and determine that it is a global solution in our case.
(3)
We can check $\widehat h$ to see if these equations are solvable.
We will do this below and actually solve the simplified SDE and the diffeomorphism
in the extended Heston case.
(4)
It is shown in \citet{Kouritzin/Remillard:2015} that (\ref{Bracket1}) is also
necessary if we want to have such local solutions for all initial
random variables $X_0$.

In our Heston case $\overline X=\left(\overline Y,\overline Z \right)'$
and $\widehat X=\widehat S^c$ and we can use Theorem \ref{KoRe} to obtain:
\begin{thm}\label{ReduceHeston}
Suppose $(W^1,W^2)'$ is a standard $\mathbb R^2$-valued Brownian motion and
$\left(\overline Y_t,\overline Z_t,\widehat S^c_t \right)'$
is the strong solution to:
\begin{eqnarray}
d\left[\!\begin{array}{c}\overline Y_t\\\overline Z_t\end{array}\!\right]&\!\!\! =&\!\!\! \left[\!\begin{array}{c}-\frac{\varrho}2\overline Y_t\\
-\frac{\varrho}2\overline Z_t\end{array}\!\right] dt+d\left[\!\begin{array}{c}W^1_{t}\\W^2_{t}\end{array}\!\right],\\
d\widehat S^c_t
&\!\!\!=&\!\!\!\widehat S^c_t\!
\left[\mu -\frac{\kappa\rho}2+\left[\frac{\kappa\rho\varrho}4-\frac{\kappa^2\rho^2}8\right]\left\{\overline Y_t^2+\overline Z_t^2\right\}
\right]
 dt.
\end{eqnarray}
Then,  $\left[\!\begin{array}{c}Y_t\\ Z_t\\ S^c_t
\end{array}\!\right]=\Lambda^{-1}\left(\begin{array}{c}\overline Y_t\\\overline Z_t\\\widehat S^c_t
\end{array}\right)$ with $\left(\begin{array}{c}W^1_t\\W^2_t
\end{array}\right)$ satisfies (\ref{ExtendHesStrat}) (or equivalently (\ref{ExtendHes},\ref{newBMs})), where
\begin{equation}\label{fuldiffeo}
\Lambda(x)=\left[\!\begin{array}{c}\frac2{\kappa}\,x_1\\\frac2{\kappa}\,x_2\\
x_3\exp\left(-\frac\rho\kappa (x_1^2+x^2_2)\right)
\end{array}\!\right],\
\Lambda^{-1}(x)=\left[\!\begin{array}{c}\frac{\kappa}2\,x_1\\\frac{\kappa}2\,x_2\\x_3\exp\left(\rho\frac{\kappa}4
(x_1^2+x^2_2)\right)
\end{array}\!\right],
\end{equation}
is a $\mathcal C^2$-diffeomorphism on $\mathbb R\times \mathbb R\times (0,\infty)$.
\end{thm}
\begin{RM}
	\emph{
We do not need Condition (C) for this theorem nor even for the solution of
price $S$ in terms of $V$ below.
We only need this condition to express the volatility in terms of the
sums of squares of independent Ornstein-Uhlenbeck processes.
}
\end{RM}
\begin{RM}
	\emph{
We only really care that we have a solution for the last rows of (\ref{ExtendHes},\ref{newBMs})
but we have to solve for all rows and then later throw away the unnecessary ones.
}
\end{RM}
\begin{RM}
\emph{
$\overline Y$ and $\overline Z$ are independent Ornstein-Uhlenbeck
processes while
$\widehat S^c$ just solves a linear ordinary
differential equation (with coefficients depending upon the random processes
$\overline Y,\overline Z$).
Hence, simulation and calculation is made easy by the explicit form of the
diffeomorphism and its inverse.
Notice that $\widehat S^c $ has finite variation while
$S^c$ does not.
The explanation for this is that the diffeomorphism $\Lambda^{-1}$ brings $\overline Y$ and $\overline Z$
into the solution for $S^c$ and thereby handles the quadratic variation.
}
\end{RM}
\proof
The idea is to find the diffeomorphisms $\Lambda_1,\Lambda_2$ in Theorem \ref{KoRe}.
Solving $\frac{d}{dt}\theta(t;x)=\sigma_1(\theta(t;x))$ with $\sigma$
as in (\ref{ExtendHesSigma}) leads to
\begin{eqnarray}\label{ThetaSigma1}
	\frac{d}{dt}\theta(t;x)=\!\left[\!\begin{array}{c}\frac{\kappa}2\\0\\
\rho\, \theta_1(t;x)\theta_3(t;x)\end{array}\!\right]
\ \text{subject to }\theta(0;x)=\left[\!\begin{array}{c}0\\x_2\\
x_3\end{array}\!\right]\!,
\end{eqnarray}
and we find that
\(
\theta_1(t;x)=\frac{\kappa}2t;\ \theta_2(t;x)=x_2;\ \)
\(
	\theta_3(t;x)=x_3\exp
\left(\frac{\rho\kappa}4\,t^2\right)
\).
Substituting $t=x_1$ in, we have that
\begin{eqnarray}\label{PsiSigma1}
\Lambda_1^{-1}(x)=\left[\begin{array}{c}\frac{\kappa}2\,x_1\\x_2\\
x_3\exp \left(\frac{\rho\kappa}4\,x_1^2\right) \end{array}\right],
\end{eqnarray}
which has inverse
\begin{eqnarray}\label{LambdaSigma1}
\Lambda_1(y)=\left[\begin{array}{c}\frac2{\kappa}\,y_1\\y_2\\
y_3\exp \left(-\frac{\rho}{\kappa}\,y_1^2\right)\end{array}\right].
\end{eqnarray}
Next, it follows that
\begin{equation}\label{LambdaSigma1}
\!\!\!\!\!\nabla\!\Lambda_1(y)=
\left[\!\begin{array}{ccc}
\frac2{\kappa}&0&0\\ 0&1&0\\
-2\frac{\rho}{\kappa}\,y_1y_3\exp \left(-\frac{\rho}{\kappa}\,y_1^2\right)& 
0
&\exp \left(-\frac{\rho}{\kappa}\,y_1^2\right)
\end{array}\right]
\end{equation}
so $\widehat\sigma_1(x)=\{\nabla\Lambda_1\sigma_1\}(\Lambda_1^{-1}(x))=e_1$ and we have
found our first diffeomorphism in Theorem \ref{KoRe}.
To find the second diffeomorphism, we set
\begin{equation}\label{Sigmahat}
\alpha_2(x)=\{\nabla\Lambda_1\sigma_2\}(\Lambda_1^{-1}(x))=\left[\begin{array}{c}
0\\
\frac{\kappa}2\\
\rho\ x_2 x_3\end{array}\right].
\end{equation}
Then, solving $\frac{d}{dt}\theta(t;x)=\alpha_2(\theta(t;x))$ leads to
\begin{eqnarray}\label{ThetaSigma1}
\frac{d}{dt}\theta(t;x)=\!\left[\!\begin{array}{c}0\\\frac{\kappa}2\\
\rho\,\theta_2(t;x)\,\theta_3(t;x)\end{array}\!\right]
\ \text{s.t.\ }\theta(0;x)=\left[\!\begin{array}{c}x_1\\0\\
x_3\end{array}\!\right]\!,
\end{eqnarray}
and we find that
\(
\theta_1(t;x)=x_1;\ \theta_2(t;x)=\frac{\kappa}2t;\ \)
\(
\theta_3(t;x)=x_3\exp\left(\frac{\rho\kappa}{4}\,t^2\right).
\)
Substituting $t=x_2$ in and taking the inverse, we have that
\begin{equation}
\Lambda^{-1}_2(x)=\left[\begin{array}{c}x_1\\\frac{\kappa}2x_2\\
x_3\exp \left(\frac{\rho\kappa}{4}\,x_2^2\right)\end{array}\right],\
\label{LambdaSigma2}
\Lambda_2(y)=\left[\begin{array}{c}y_1\\\frac2{\kappa}\,y_2\\
y_3\exp \left(-\frac{\rho}{\kappa}\,y_2^2\right) \end{array}\right].
\end{equation}
Next, it follows that
\begin{eqnarray}\label{GradLambdaSigma2}
\nabla\Lambda_2(y)=\left[\begin{array}{ccc}
1&0&0\\
0&\frac2{\kappa}&0\\
0& 
-2\frac\rho{\kappa}\,y_2\,y_3\exp \left(-\frac{\rho}{\kappa}\,y_2^2\right)
&\exp \left(-\frac{\rho}{\kappa}\,y_2^2\right)
\end{array}\right]
\end{eqnarray}
so
\(
\widehat\sigma_2(x)=\{\nabla\Lambda_2\alpha_2\}(\Lambda_2^{-1}(x))=e_2 
\)
and we indeed have our second homeomorphism in Theorem \ref{KoRe}.
Now, we find $\Lambda=\Lambda_2\circ\Lambda_1$ gives the diffeomorphism
in (\ref{fuldiffeo}) and 
\begin{equation}\label{GradLambda}
\!\!\nabla\Lambda (y)=
\!\left[\!\begin{array}{ccc}
\frac2{\kappa}&0&0\\
0&\frac2{\kappa}&0\\
\frac{-2\frac\rho{\kappa}\,y_1y_3}{\exp\left(\frac\rho\kappa (y_1^2+y^2_2)\right)}&
\frac{-2\frac\rho{\kappa}\,y_2y_3}{\exp\left(\frac\rho\kappa (y_1^2+y^2_2)\right)}&
\frac1{\exp\left(\frac\rho\kappa (y_1^2+y^2_2)\right)} \end{array}\right]
\end{equation}
so $\widehat h(x)\doteq (\nabla \Lambda)h\circ\Lambda^{-1}(x)$ in Theorem \ref{KoRe} satisfies
\begin{eqnarray}\label{hhat}
\widehat h(x)=\!\left[\!\begin{array}{c}
-\frac{\varrho}2 x_1\\
-\frac{\varrho}2 x_2\\
x_3\!\left[\mu -\frac{\kappa\rho}2+\left[\frac{\kappa\rho\varrho}4-\frac{\kappa^2\rho^2}8\right]\left\{x_1^2+x_2^2\right\}
\right]
\end{array}\!\!\right].
\end{eqnarray}
\qed 

\subsubsection{Finishing Proof of Theorem \ref{Theorem1} by Solving Equations in case $n=2$}

The solution for $\left(\overline Y_t,\overline Z_t,\widehat S^c_t \right)'$ 
in Theorem \ref{ReduceHeston} is:
$\overline Y_t=\int_0^t e^{-\frac{\varrho}2 (t-u)}dW^1_u+e^{-\frac{\varrho}2 t}\overline Y_0$,
$\overline Z_t=\int_0^t e^{-\frac{\varrho}2 (t-u)}dW^2_u+e^{-\frac{\varrho}2 t}\overline Z_0$ (with $\overline Y_0^2+\overline Z_0^2=\frac4{\kappa^2}  V_0$
to be consistent with (\ref{CIRModel},\ref{BasicYZ})),
and
\begin{equation}\label{Stildeclosed}
\widehat S^c_t=\widehat S^c_0
\exp\bigg(
\left[\mu -\frac{\kappa\rho}2\right] t
+\left[\frac{\kappa\rho\varrho}4-\frac{\kappa^2\rho^2}8\right]
\int_0^t
\left\{\overline Y_s^2+\overline Z_s^2\right\} ds
\bigg).
\end{equation}
Moreover, it follows by (\ref{fuldiffeo}) and (\ref{BasicYZ}) that
\begin{equation}
S^c_t=\widehat S^c_t
\exp\left(\frac{\rho\kappa}4 (\overline Y_t^2+\overline Z_t^2)\!\right)
=\widehat S_t^c\exp\left(\frac\rho\kappa (Y_t^2+Z_t^2)\!\right)
=\widehat S_t^c\exp\left(\frac\rho\kappa  V_t\!\right)
\end{equation}
and it follows by (\ref{Stildeclosed}), Theorem \ref{ReduceHeston}, (\ref{fuldiffeo}) and substitution
that
\begin{eqnarray}\label{Snottildeclosed}
\!\!\!\!\!\!S^c_t&\!\!\!=&\!\!\! S^c_0
\exp\!\bigg(
\!\left[\mu-\frac{\kappa\rho}2\right] t
+\left[\frac{\kappa\rho\varrho}4-\frac{\kappa^2\rho^2}8\right]
\int_0^t\!
\left\{\overline Y_s^2+\overline Z_s^2\right\} ds+\frac\rho\kappa ( V_t- V_0)\!
\bigg)\\\nonumber
\!\!&\!\!\!=&\!\!\! S^c_0
\exp\!\bigg(
\!\left[\mu-\frac{\kappa\rho}2\right] t
+\left[\frac{\rho\varrho}\kappa-\frac{\rho^2}2\right]
\int_0^t V_s ds+\frac\rho\kappa ( V_t- V_0)\!
\bigg).
\end{eqnarray}
We also get a solution for the simplified Heston (\ref{SimpHeston}) by
computing 
\begin{equation}\label{SiNew}
S^i_t=\exp\left(\sqrt{1-\rho^2}\int_0^t  V_s^\frac12 dB_s-\frac{1-\rho^2}2\int_0^t  V_sds\right)
\end{equation} 
and then multiplying $ S_t=S^c_tS^i_t$ to get (\ref{ExplicitSt})
of Theorem \ref{Theorem1} in the case $n=2$. \qed

\subsubsection{Case $n\ne2$}

Insomuch as the guess and check proof of Theorem \ref{Theorem1} is as simple as It\^{o}'s
formula, our real goal here is to motivate how this solution was actually arrived
at and how weak solutions for other models might be found.
With this easy Ito lemma test, a formal proof along these lines is less important.
Hence, we have given all the steps just in the case $n=2$ and we will just explain the
differences required for the case $n\ne 2$ instead of going through the formal proof
with these methods.

The price splitting was already done in general. 
There is no change there.

For the volatility in the case $n\in\{1,3,4,...\}$, we start with $n$ independent
standard Brownian motions $W^1,...,W^n$ and follow Subsection \ref{Volatility2}.
The differences are: 
We replace $Y,Z$ with $\{Y_t^i=\frac{\kappa}2\int_0^t e^{-\frac{\varrho}2 (t-u)}dW^i_u+e^{-\frac{\varrho}2t}Y^i_0\}_{i=1}^n$
and set
\begin{equation}\label{betan2}
\widehat\beta_t=\sum_{i=1}^n \int_0^t\frac{Y_u^i}{\sqrt{\sum_{j=1}^n(Y_u^j)^2}}dW^i_u
\end{equation}
to find that $V=\sum\limits_{i=1}^n (Y^i)^2$ satisfies (\ref{CIRModel}) when $\nu=\frac{n\kappa^2}4$ (and $V_0=\sum\limits_{i=1}^n(Y^i_0)^2$).

For the extended price formulation when $n\in\{1,3,4,...\}$, we set
\begin{equation}\label{ExtendHesSigman}
\sigma(y_1,...,y_n,s)=\left[\begin{array}{ccccc}
\frac{\kappa}{2}&0&0&\cdots&0\\
0&\frac{\kappa}{2}&0&\cdots&0\\
\vdots&\vdots&\ddots&\vdots&\vdots\\
0&0&\cdots&\frac{\kappa}{2}&0\\
0&0&\cdots&0&\frac{\kappa}{2}\\
s\rho\,y_1& s\rho\, y_2
&\cdots&s\rho\,y_{n-1}& s\rho\, y_n
\end{array}\right]
\end{equation}
and find $\nabla\sigma_i\sigma_j=(0,...,0,s\rho^2y_iy_j)'$ for $i\ne j$ so (\ref{Bracket1}) clearly holds.
(For clarity, $\sigma=(\frac\kappa2,s\rho y_1)'$ when $n=1$.)
Now, define a new SDE of the form:
\begin{equation}\label{ExtendHesan}
d\!\left[\begin{array}{c}Y_t^1\\\vdots\\Y^n_t\\S^c_{t}\end{array}\right]
=\left[\begin{array}{c}-\frac{\varrho}2 Y^1_t\\\vdots\\-\frac{\varrho}2 Y^n_t\\\mu  S^c_t
\end{array}\right]dt+\!\sigma(Y_t^1,...,Y^n_t,S^c_t)\left[\begin{array}{c}dW^1_{t}\\\vdots\\dW^n_{t}\end{array}\right].
\end{equation}
This equation has a unique strong solution 
and  it can be rewritten by postmultiplying $\sigma$ by $OO^{-1}$, where
\begin{equation}\label{ExtendHesO}
O=\left[\begin{array}{ccccc}
\frac{Y_t^n}{\sqrt{V_t}}&0&\cdots&0&\frac{Y_t^1}{\sqrt{V_t}}\\
0&\frac{Y_t^n}{\sqrt{V_t}}&\cdots&0&\frac{Y_t^2}{\sqrt{V_t}}\\
\vdots&\vdots&\ddots&\vdots&\vdots\\
		0&0&\cdots&\frac{Y_t^n}{\sqrt{V_t}}&\frac{Y_t^{n-1}}{\sqrt{V_t}}\\
-\frac{Y_t^1}{\sqrt{V_t}}&-\frac{Y_t^2}{\sqrt{V_t}}&\cdots&-\frac{Y_t^{n-1}}{\sqrt{V_t}}&\frac{Y_t^n}{\sqrt{V_t}}
\end{array}\right],
\end{equation}
and (abusing notation by letting $Y_i=Y^i_t$)
\begin{equation}\label{ExtendHesO1}
O^{-1}=\left[\begin{array}{cccccc}
\frac{Y_2^2+\cdots+Y_n^2}{Y_n\sqrt{V_t}}&-\frac{Y_1Y_2}{Y_n\sqrt{V_t}}&-\frac{Y_1Y_3}{Y_n\sqrt{V_t}}&\cdots&-\frac{Y_1Y_{n-1}}{Y_n\sqrt{V_t}}&-\frac{Y_1}{\sqrt{V_t}}\\
-\frac{Y_1Y_2}{Y_n\sqrt{V_t}}&\frac{Y_1^2+Y_3^2+\cdots+Y_n^2}{Y_n\sqrt{V_t}}&-\frac{Y_2Y_3}{Y_n\sqrt{V_t}}&\cdots&-\frac{Y_2Y_{n-1}}{Y_n\sqrt{V_t}}&-\frac{Y_2}{\sqrt{V_t}}\\
\vdots&\vdots&\vdots&\ddots&\vdots&\vdots\\
-\frac{Y_1Y_{n-1}}{Y_n\sqrt{V_t}}&-\frac{Y_2Y_{n-1}}{Y_n\sqrt{V_t}}&-\frac{Y_3Y_{n-1}}{Y_n\sqrt{V_t}}&\cdots&\frac{Y_1^2+\cdots+Y_{n-2}^2+Y_n^2}{Y_n\sqrt{V_t}}&-\frac{Y_{n-1}}{\sqrt{V_t}}\\
\frac{Y_1}{\sqrt{V_t}}&\frac{Y_2}{\sqrt{V_t}}&\frac{Y_3}{\sqrt{V_t}}&\cdots&\frac{Y_{n-1}}{\sqrt{V_t}}&\frac{Y_n}{\sqrt{V_t}}
\end{array}\right],
\end{equation}
as:
\begin{equation}\label{ExtendHesn}
d\!\left[\begin{array}{c}Y^1_t\\\vdots\\Y^n_t\\S^c_{t}\end{array}\right]
\!=\!\left[\begin{array}{c}-\frac{\varrho\,Y^1_t}2\\\vdots \\-\frac{\varrho\,Y^n_t}2 \\\mu  S^c_t\end{array}\right]dt+\!\left[\begin{array}{ccccc}
\frac{\kappa}{2}\frac{Y_t^n}{\sqrt{V_t}}&0&\cdots&0&\frac{\kappa}{2}\frac{Y_t^1}{\sqrt{V_t}}\\
\vdots&\vdots&\ddots&\vdots&\vdots\\
0&0&\cdots&\frac{\kappa}{2}\frac{Y_t^n}{\sqrt{V_t}}&\frac{\kappa}{2}\frac{Y_t^{n-1}}{\sqrt{V_t}}\\
-\frac{\kappa}{2}\frac{Y_t^1}{\sqrt{V_t}}&-\frac{\kappa}{2}\frac{Y_t^2}{\sqrt{V_t}}&\cdots&-\frac{\kappa}{2}\frac{Y_t^{n-1}}{\sqrt{V_t}}&\frac{\kappa}{2}\frac{Y_t^{n}}{\sqrt{V_t}}\\
0 &0 &\cdots &0 &\rho S^c_{t} V_{t}^{\frac{1}{2}}\end{array}\right]
\left[\begin{array}{c}dA^1_{t}\\\vdots\\dA^{n-1}_{t}\\d\widehat \beta_{t}\end{array}\right],
\end{equation}
where $(A^1,...,A^{n-1},\widehat\beta)'=O^{-1}(W^1,...,W^n)'$ so $\widehat\beta$ does satisfy
(\ref{betan2}).
This extended Heston solution (\ref{ExtendHesan}) can also be written in Fisk-Stratonovich
form as
\begin{equation}\label{ExtendHesanFisk}
d\!\left[\begin{array}{c}Y_t^1\\\vdots\\Y^n_t\\S^c_{t}\end{array}\right]
=\left[\begin{array}{c}-\frac{\varrho}2 Y^1_t\\\vdots\\-\frac{\varrho}2 Y^n_t\\\left(\mu-\frac{n\kappa\rho}4\right)  S^c_t
-S^c_t\rho^2\frac{(Y_t^1)^2+\cdots+(Y^n_t)^2}2		
\end{array}\right]dt+\!\sigma(Y_t^1,...,Y^n_t,S^c_t)\bullet\left[\begin{array}{c}dW^1_{t}\\\vdots\\dW^n_{t}\end{array}\right],
\end{equation}
from which we can apply Theorem 2 of \cite{Kouritzin/Remillard:2015} 
(knowing (\ref{Bracket1}) holds) in the case $p=n+1$ and $d=r=n$
to find (\ref{ExtendHesanFisk}) has a strong solution up to some stopping time
$\tau>0$ if and only if
\begin{eqnarray}\label{starst1}
d\left[\!\begin{array}{c}\overline Y^1_t\\\vdots\\\overline Y^n_t\end{array}\!\right]&\!\!\! =&\!\!\! \left[\!\begin{array}{c}-\frac{\varrho}2\overline Y^1_t\\
\vdots\\-\frac{\varrho}2\overline Y^n_t\end{array}\!\right] dt+d\left[\!\begin{array}{c}W^1_{t}\\\vdots\\W^n_{t}\end{array}\!\right],\\
d\widehat S^c_t\label{starst2}
&\!\!\!=&\!\!\!\widehat S^c_t\!
\left[\mu -\frac{n\kappa\rho}4+\left[\frac{\kappa\rho\varrho}4-\frac{\kappa^2\rho^2}8\right]\left\{\left(\overline Y^1_t\right)^2+\cdots+\left(\overline Y^n_t\right)^2\right\}
\right]
 dt
\end{eqnarray}
does.
Moreover, the solutions to (\ref{ExtendHesanFisk}) and (\ref{starst1},\ref{starst2})
satisfy
\begin{equation}
\left[\!\begin{array}{c}Y^1_t\\\vdots\\Y^n_t\\ S^c_t\end{array}\!\right]
=\Lambda^{-1}\left(\!\begin{array}{c}\overline Y^1_t\\\vdots\\\overline Y^n_t\\\widehat S^c_t\end{array}\!\right),
\end{equation}
where $\mathcal C^2$-diffeomorphism $\Lambda$ is given by
\begin{equation}\label{fuldiffeon}
\Lambda(x)=\left[\!\begin{array}{c}\frac2{\kappa}\,x_1\\\vdots\\\frac2{\kappa}\,x_n\\
x_{n+1}\exp\left(-\frac\rho\kappa (x_1^2+\cdots+x^2_n)\right)
\end{array}\!\right],\
\Lambda^{-1}(x)=\left[\!\begin{array}{c}\frac{\kappa}2\,x_1\\\vdots\\\frac{\kappa}2\,x_n\\x_{n+1}\exp\left(\rho\frac{\kappa}4
(x_1^2+\cdots+x^2_n)\right)
\end{array}\!\right].
\end{equation}
The solution to (\ref{starst1},\ref{starst2}) is then
\begin{equation}\label{Stildeclosedn}
\overline Y^i_t=\int_0^t e^{-\frac{\varrho}2 (t-u)}dW^i_u+e^{-\frac{\varrho}2 t}\overline Y^i_0,\
i=1,...,n\ \text{and}
\end{equation}
\begin{equation}\label{Stildeclosedn}
\!\!\widehat S^c_t=\widehat S^c_0
\exp\bigg(
\left[\mu -\frac{n\kappa\rho}4\right] t
+\left[\frac{\kappa\rho\varrho}4-\frac{\kappa^2\rho^2}8\right]
\int_0^t
\left\{\!\left(\overline Y^1_s\right)^2+\cdots+\left(\overline Y^n_s\right)^2\!\right\} ds
\bigg)
\end{equation}
from which it follows using (\ref{fuldiffeon}) that
\begin{equation}\label{Stildeclosedn}
S^c_t=S^c_0
\exp\bigg(
\left[\mu -\frac{n\kappa\rho}4\right] t
+\left[\frac{\rho\varrho}\kappa-\frac{\rho^2}2\right]
\int_0^t V_s ds+\frac{\rho}{\kappa}(V_t-V_0)
\bigg)
\end{equation}
with 
$\displaystyle V_t=\frac{\kappa^2}4\left\{\left(\overline Y^1_t\right)^2+\cdots+\left(\overline Y^n_t\right)^2\right\}$.
The result follows by multiplying $S_t =S^i_t S^c_t$ and It\^{o}'s formula.  $\square$

\subsection{Proof of Theorem \ref{Theorem2}}
We follow ideas that could be used to prove Girsanov's theorem noting that
the solutions are weak so martinagle problems not SDEs are the correct tools
and $L$ is the form for easy simulation not for direct change of measure.
By Theorem \ref{Theorem1}, $(\widehat S,\widehat V)$, defined
in (\ref{ExplicitStful},\ref{ExplicitVtful}) satisfies
the Heston model with parameters $\nu_\kappa,\mu_\kappa$ defined in (\ref{numuk}).
Hence, by (\ref{LHestdef}) 
\begin{eqnarray}
M_t(f)&\!\!=&\!\!
f(\widehat S_t,\widehat V_t)-\int_0^t \mu_\kappa \widehat S_u\,\partial_{s}f(\widehat S_u,\widehat V_u)+(\nu_\kappa-\varrho \widehat V_u) \partial_{v}f(\widehat S_u,\widehat V_u)  \label{LHestdef1}\\\nonumber 
&\!\!+&\!\!\frac{1}{2}\widehat S_u^2\widehat V_u\,\partial^2_sf(\widehat S_u,\widehat V_u)
+\rho\kappa \widehat S_u\widehat V_u\,\partial_{s}\partial _{v}f(\widehat S_u,\widehat V_u)
+\frac{1}{2}\kappa^2\widehat V_u\,\partial^2_vf(\widehat S_u,\widehat V_u) du
\end{eqnarray}
(for $f\in \mathcal S(\mathbb R^2)$, the rapidly decreasing functions) has the following $P$-martingale representation
\begin{eqnarray}\label{ExpHesMart}
M_t(f)&\!\!\!=&\!\!\!\!\int_0^{t}\! [\kappa \partial_v f(\widehat S_u,\widehat V_u)
+\rho \widehat S_{u}\partial_s f(\widehat S_u,\widehat V_u)]\widehat V_{u}^{\frac{1}{2}}d\widehat \beta_u\\\nonumber
&\!\!\!+&\!\!\!\!\int_0^{t}\!\sqrt{1-\rho^2}\widehat 
S_{u}\partial_s f(\widehat S_u,\widehat V_u)\widehat V_{u}^{\frac{1}{2}}dB_u\ \text{ with } 
\widehat \beta_t=\sum_{i=1}^n \int_0^t\frac{Y_u^i}{\sqrt{\sum_{j=1}^n(Y_u^j)^2}}dW^i_u.
\end{eqnarray}
Separately, it follows by It\^{o}'s formula and (\ref{SimpHeston}) that
\begin{equation}
\ln(\widehat V_t)-\ln(\widehat V_0)=\int_0^t \frac{\nu_\kappa-\varrho \widehat V_s}{\widehat V_s}ds +\int_0^t \frac{\kappa}{\widehat V_s^\frac12}d\widehat\beta_s-\frac12\int_0^t \frac{\kappa^2}{\widehat V_s}ds
\end{equation}
so, using (\ref{numuk}), (\ref{L2way}) is equivalent to
\begin{equation}\label{L1way}
\widehat L_t=\exp\left\{\int_0^t \frac{\nu-\nu_\kappa}{\kappa \widehat V_s^\frac12}d\widehat\beta_s
-\frac12\int_0^t \frac{|\nu-\nu_\kappa|^2}{\kappa^2 \widehat V_s}ds
\right\}.
\end{equation}
It follows from (\ref{L1way}) and the Novikov condition that
$t\rightarrow \widehat L^{\eta_\varepsilon}_t\doteq \widehat L_{\eta_\varepsilon\wedge t}$ is an
$L^r$-martingale for any $r>0$.
This fact will be used in the development below and to conclude $m_t(f)$ is a martingale versus just a local martingale.
Next, it follows by (\ref{ExpHesMart}), It\^{o}'s formula, (\ref{numuk}) and the fact 
\(
d\widehat L_t=\widehat L_t\frac{\nu-\nu_\kappa}{\kappa }\widehat V_t^{-\frac12}d\widehat\beta_t
\)
(by (\ref{L1way})) that the quadratic covariance satisfies
\begin{eqnarray}\label{LgfComp}
\!\!\!\!\!\!\!\!\!\![\widehat L^{\eta_\varepsilon},f(\widehat S,\widehat V)]_t&\!\!\!\!=&\!\!\!\!
\int_0^{t\wedge\eta_\varepsilon}\! \widehat L_u^{\eta_\varepsilon}
\frac{\nu-\nu_\kappa}{\kappa }\widehat V_u^{-\frac12}\!\left[\kappa\partial_vf(\widehat S_u,\widehat V_u)
+\rho \widehat S_{u}\partial_s f(\widehat S_u,\widehat V_u)
\right]\! \widehat V_u^\frac12du\\\nonumber
&\!\!\!=&\!\!\!\int_0^{t\wedge\eta_\varepsilon} \widehat L^{\eta_\varepsilon}_u\left[(\nu-\nu_\kappa)\partial_vf(\widehat S_u,\widehat V_u)
+(\mu-\mu_\kappa) \widehat S_{u}\partial_s f(\widehat S_u,\widehat V_u)
\right] du.
\end{eqnarray}
Now, it follows by (\ref{LHestdef1},\ref{LgfComp}) and integration by parts that
\begin{eqnarray}
\!m_t(f)&\!\!\!=&\!\!\!
\widehat L_t^{\eta_\varepsilon}f(\widehat S_t,\widehat V_t)-\!\int_0^{t\wedge\eta_\varepsilon}\!\widehat  L_u^{\eta_\varepsilon}\big[\mu \widehat S_u\,\partial_{s}f(\widehat S_u,\widehat V_u)+(\nu-\varrho \widehat V_u) \partial_{v}f(\widehat S_u,\widehat V_u) \big] du \label{IntPartHestdef}
\\\nonumber 
&\!\!\!-&\!\!\!\int_{t\wedge\eta_\varepsilon}^t \widehat L^{\eta_\varepsilon}_u\big[\mu_\kappa \widehat S_u\,\partial_{s}f(\widehat S_u,\widehat V_u)+(\nu_\kappa-\varrho \widehat V_u) \partial_{v}f(\widehat S_u,\widehat V_u) \big] du\\\nonumber 
&\!\!\!-&\!\!\!\int_0^t\widehat  L_u^{\eta_\varepsilon}\bigg[ \frac{1}{2}\widehat S_u^2\widehat V_u\,\partial^2_sf(\widehat S_u,\widehat V_u)
+\rho\kappa \widehat S_u\widehat V_u\,\partial_{s}\partial _{v}f(\widehat S_u,\widehat V_u)
+\frac{1}{2}\kappa^2\widehat V_u\,\partial^2_vf(\widehat S_u,\widehat V_u)\bigg] du
\end{eqnarray}
is a local martingale, which by (\ref{ExpHesMart}) has form
\begin{eqnarray}\label{WeightHesMart}
\!\!\!\!m_t(f)&\!\!\!=&\!\!\!\!\int_0^{t}\! \widehat L^{\eta_\varepsilon}_u[\kappa \partial_v f(\widehat S_u,\widehat V_u)
+\rho \widehat S_{u}\partial_s f(\widehat S_u,\widehat V_u)
+\frac{\nu-\nu_\kappa}{\kappa \widehat V_u}f(\widehat S_u,\widehat V_u)]\widehat V_{u}^{\frac{1}{2}}d\widehat \beta_u\\\nonumber
&\!\!\!+&\!\!\!\!\int_0^{t}\!\widehat L^{\eta_\varepsilon}_u\sqrt{1-\rho^2}\widehat 
S_{u}\partial_s f(\widehat S_u,\widehat V_u)\widehat V_{u}^{\frac{1}{2}}dB_u.
\end{eqnarray}
(Since we have used other randomness to create the $\{Y^i\}_{i=1}^n$ we can not conclude that $m_t(f)$ is 
adapted to the filtration generated by $\beta,B$ but it is adapted to
the filtration created by $B,W^1,...,W^n$.)

Now, $\widehat L_{t}^{\eta_\varepsilon}$ and $m_{t}^{\eta_\varepsilon}(f)\doteq m_{t\wedge\eta_\varepsilon}(f)$ are martingales so one has by (\ref{IntPartHestdef}) and Fubini's theorem that
\begin{eqnarray}\label{PtildeJointEv}
&\!\!\!\!&\!\!\!\!\widehat E\left[\left(f(\widehat S_{t_{n+1}},\widehat V_{t_{n+1}})-f(\widehat S_{t_{n}},\widehat V_{t_{n}})
-\int_{t_{n}}^{t_{n+1}}\!\!A_uf(\widehat S_u,\widehat V_u)du\right)
\prod_{k=1}^nh_k(\widehat S_{t_k},\widehat V_{t_k})\right]\\\nonumber
&\!\!\!\!=&\!\!\!\!E\left[\widehat L^{\eta_\varepsilon}_T\left(f(\widehat S_{t_{n+1}},\widehat V_{t_{n+1}})-f(\widehat S_{t_{n}},\widehat V_{t_{n}})
-\int_{t_{n}}^{t_{n+1}}\!\!A_uf(\widehat S_u,\widehat V_u)du\right)
\prod_{k=1}^nh_k(\widehat S_{t_k},\widehat V_{t_k})\right]\\\nonumber
&\!\!\!\!=&\!\!\!\!E\left[\left(m_{t_{n+1}}(f)-m_{t_{n}}(f)\right)
\prod_{k=1}^nh_k(\widehat S_{t_k},\widehat V_{t_k})\right]=0,
\end{eqnarray}
for all $0\le t_1< t_2<\cdots<t_n<t_{n+1}$, $f\in \mathcal S(\mathbb R^2)$ and $h_1,...,h_n\in B(\mathbb R^2)$ (the bounded, measurables),
where
\begin{eqnarray}
A_uf(s,v)&\!\!\!=&\!\!\![\mu s \partial_sf(s,v)+(\nu-\varrho v) \partial_vf(s,v)]1_{[0,\eta_\varepsilon]}(u)
\\\nonumber
&\!\!\!+&\!\!\![\mu_\kappa s \partial_sf(s,v)+(\nu_\kappa-\varrho v) \partial_vf(s,v)]1_{[\eta_\varepsilon,T]}(u)\\\nonumber
&\!\!\!+&\!\!\!\frac12 s^2 v \partial^2_sf(s,v)+\rho\kappa\partial_v\partial_sf(s,v)+\frac{\kappa^2}2\partial^2_vf(s,v).
\end{eqnarray}
Now, it follows by the argument on page 174 of \citet{Ethier/Kurtz:1986} that 
$(S,V)$ satisfies the $A_u$-martingale problem with respect to $\widehat P\ \square$.

\bibliographystyle{apalike}

\bibliography{Sim12}
\bigskip
	
\end{document}